\documentclass[aps,prx,onecolumn,longbibliography,amsmath,amssymb,nofootinbib,superscriptaddress]{revtex4-2}

\usepackage{graphicx}
\usepackage[T1]{fontenc}
\usepackage[utf8]{inputenc}
\usepackage{xcolor}
\usepackage[colorlinks=true,allcolors=blue]{hyperref}

\makeatletter
\renewcommand{\p@subsection}{}
\makeatother

\begin{document}

\title{Perspective on AI and quantum open-source software}

\author{Changhao Li}
\thanks{changhao@unitary.foundation}
\affiliation{Unitary Foundation, San Francisco, CA, USA}
\author{Bradley A. Chase}
\affiliation{Unitary Foundation, San Francisco, CA, USA}
\author{Veena Vijayakumar}
\affiliation{Unitary Foundation, San Francisco, CA, USA}
\author{Mathys Rennela}
\affiliation{Unitary Foundation, San Francisco, CA, USA}
\author{Alessandro~Cosentino}
\affiliation{Unitary Foundation, San Francisco, CA, USA}
\author{Naixu Guo}
\affiliation{Centre for Quantum Technologies, National University of Singapore, 117543, Singapore}
\author{Sebastian Hassinger}
\affiliation{Unitary Foundation, San Francisco, CA, USA}
\author{Farrokh Labib}
\affiliation{Unitary Foundation, San Francisco, CA, USA}
\author{Seun Omonije}
\affiliation{Unitary Foundation, San Francisco, CA, USA}
\author{Patrick Rebentrost}
\affiliation{Centre for Quantum Technologies, National University of Singapore, 117543, Singapore}
\author{Vincent Russo}
\affiliation{Unitary Foundation, San Francisco, CA, USA}
\author{Nathan Shammah}
\affiliation{Unitary Foundation, San Francisco, CA, USA}
\author{William J. Zeng}
\affiliation{Unitary Foundation, San Francisco, CA, USA}
\author{Ben Castanon}
\affiliation{Unitary Foundation, San Francisco, CA, USA}


\begin{abstract}
Quantum open-source software  provides essential infrastructure for quantum science, and artificial intelligence (AI) is reshaping its capabilities and transforming its development. However, verifying quantum software at scale remains difficult, and AI-generated contributions can increase the workload of maintainers. In this Perspective, we examine verification as a central requirement for reliable AI-assisted quantum software development and research. We analyze public repositories, records covering 110 repositories from five annual quantum open-source hackathons that we organize, and responses from more than 1600 participants in a community survey. Our analysis connects limitations in automated verification with the review burden on quantum software maintainers. 
We propose verification practices and contribution standards that require contributors to support their changes with reproducible evidence for maintainer review.
We also examine how quantum open-source software can provide infrastructure for AI-driven research whose results the community can independently verify and build upon. We maintain an open-source companion repository to support reproducibility and help projects adopt and refine these proposals. Together, these contributions help the quantum open-source software community translate advances in AI into reliable quantum research through reproducible verification.

\end{abstract}

\maketitle

\section{Introduction}
\label{sec:intro}

Quantum information science has developed rapidly in recent years with applications across computation~\cite{Ladd2010}, simulation~\cite{Georgescu2014}, sensing~\cite{Degen2017}, and communication~\cite{Wehner2018}. Progress in each of these directions depends on software, and in quantum computing that software is largely open source, from circuit construction~\cite{2405.08810,Cirq,1811.04968,McClean2020} and compilation~\cite{Sivarajah2021,BQSKit} to hardware control~\cite{Lambert2026,Puzzuoli2023,Wittler2021,Goerz2019} and error correction~\cite{2103.02202,Higgott2025,Wu2023}. Artificial intelligence (AI), including large language models (LLMs), is driving substantial advances in the capabilities of quantum open-source software (QOSS)~\cite{2411.09131,Bausch2024,2402.14396} and playing an increasingly important role in its development and maintenance~\cite{2405.19495,2503.02497,Chen2021,2310.06770}.
These tools can accelerate the discovery of new quantum methods and improve the accuracy and efficiency of existing ones.  For the QOSS community, they can broaden participation in software development by lowering technical barriers for new contributors and reducing routine implementation effort. Their output, however, becomes part of software on which the research community depends, raising questions about its continued reliability and maintainability as AI use expands. Progress in AI-assisted quantum research therefore requires advances in verification alongside advances in generation.

QOSS also provides infrastructure for AI-driven research, where verification plays a direct role in the discovery process. Open-source libraries allow agents to construct and evaluate candidate solutions, while executable verification procedures provide feedback for further exploration. In these auto-research workflows, the choice of objective and the design of the evaluator shape what agents discover~\cite{2610.23832,UnitaryAgenticLeaderboards2026}. However, success under an evaluator does not by itself establish a scientific result. Independent verification is needed to determine whether the finding supports the claimed conclusion beyond the conditions explored during the search. QOSS can make this verification available alongside the implementation, allowing other researchers to assess and extend the work. It could therefore also offer a publication format for advances in quantum research, alongside emerging paper-to-agent approaches~\cite{Miao2026Paper2Agent}.

For quantum software, verification is particularly demanding because computing reference results may itself be prohibitively expensive. For example, exact classical simulation of general quantum circuits becomes impractical beyond a few tens of qubits~\cite{Suzuki2021,Isakov2021}, while verification on quantum hardware is constrained by noise, calibration drift, and limited device access~\cite{1803.03226}.  A further limitation is that models may have incomplete or outdated knowledge of the quantum framework being used, making their outputs harder to trust without verification. Reliable AI-assisted development and auto-research therefore require explicit verification criteria and a clear statement of the scope and limitations of the supporting evidence.

What automated verification leaves unresolved falls to project maintainers, who must judge whether a contribution addresses the problem and is adequately supported. Organizers of community events report that unverified AI-generated contributions have already shifted maintainer time from mentoring toward reviewing~\cite{unitaryHACK2026}, and a similar shift has been documented across open-source software~\cite{Watanabe2026,Baltes2026,Afroz2026}. This change may also affect the training of future maintainers, who develop their skills through introductory issues, code review, and hackathons. Task completion alone does not demonstrate a contributor’s understanding of the changes, and subjective assessments of productivity may not reflect actual performance. In a controlled study, experienced developers took longer to complete tasks with AI assistance, despite believing that it had made them faster~\cite{Becker2025}. For quantum software, systematic evidence on how AI-assisted development affects maintainer workload and the training of future maintainers remains limited.

In this Perspective, we examine how verification can support reliable AI-assisted development and research across QOSS. We first survey AI applications across the quantum software stack and characterize technical constraints through analyses of public code availability, interface stability, and test circuit sizes. We then analyze repository disclosures alongside records and participant surveys from five quantum open-source hackathon events that we organize~\cite{unitaryHACK}, complemented by a broader QOSS community survey with more than 1600 respondents~\cite{QOSSsurvey2026}. Together, these sources provide evidence on verification practices and the review effort associated with AI-assisted contributions. Drawing from the hackathon events that we organize, we also assemble a collection of 417 candidate maintenance tasks from 110 repositories and characterize their verification requirements to inform future benchmarks of coding agents. Based on these analyses, we propose verification practices and contribution standards that combine automated verification with maintainer review. We further discuss how open-source research harnesses can guide agent-driven exploration toward scientifically meaningful results supported by independent verification. An open-source companion repository provides data and scripts for reproducing our analyses, alongside templates for implementing and refining these proposals~\cite{ai4qoss}.

The Perspective is organized as follows. Section~\ref{sec:landscape} reviews AI applications across the QOSS stack. Section~\ref{sec:constraints} examines technical and community constraints and discusses practices for addressing them. Section~\ref{sec:community} presents evidence from public repositories, community events, and surveys. 
Section~\ref{sec:build} presents recommendations for verification, open-source infrastructure for AI-driven research, and benchmark development alongside standards for AI-assisted contributions. Section~\ref{sec:outlook} discusses future directions for AI and quantum open-source software.

\section{AI across the quantum software stack}
\label{sec:landscape}

We organize current AI applications in QOSS into seven categories according to their functions and cover both task-specific machine-learning methods and LLM-based assistants. These applications reflect two complementary roles: AI helps develop and improve quantum software, while QOSS provides tools for AI-driven research. We use auto-research to describe workflows in which agents propose and evaluate candidate solutions, then use the results to choose subsequent steps toward a human-defined research goal. For each category, we describe the available verification methods and discuss how they support or limit practical use.

\textbf{Writing code.} 
Assistants trained on a single framework are the most mature category. 
Training and retrieval provide complementary ways to adapt LLMs to quantum programming. For example, the Qiskit Code Assistant uses a code model further trained on Qiskit code and tutorials~\cite{2405.19495}. The PennyLang study follows the second approach by providing models with relevant PennyLane code examples and descriptions from a curated dataset alongside the user's request~\cite{2503.02497}.
Agentic workflows also combine code generation with tools for executing and refining quantum programs. Examples include Pasqal's workflow connecting \texttt{Pulser} to emulation and hardware execution~\cite{2607.25834}, and qBraid's \texttt{codeq} agent for writing, running, and debugging quantum code~\cite{qBraidCodeq2026}. 
We note that existing evaluations cover quantum code generation, problem solving, and debugging~\cite{2406.14712,2604.08570,2605.27210,2511.00092,2606.07314,Pan2025}. These tasks provide evidence about specific programming capabilities, while repository maintenance also requires understanding existing interfaces, dependencies, and behavior. We discuss how community maintenance records could support evaluations of these capabilities in Section~\ref{sec:build}.

\textbf{Control and calibration.} AI-assisted control now includes LLM-based pulse design and agents that coordinate calibration experiments.
Reinforcement learning has been applied to pulse optimization and gate calibration~\cite{1803.01857,2104.14539,2105.01079,Sivak2026}, while earlier automation organized calibration procedures as dependency graphs~\cite{1803.03226}.
More recently, LLMs have been used to propose analytic control-pulse ansätze whose parameters are refined through numerical optimization~\cite{2605.26021}.
LLM agents have also coordinated superconducting-qubit calibration using laboratory knowledge and reusable experimental procedures~\cite{2412.07978,2606.22376}.
On trapped-ion platforms, agents have generated native ARTIQ control code for calibration experiments, with hardware operations screened against prescribed safety limits~\cite{2606.27231}.
Vision-language models have also been evaluated for interpreting calibration plots~\cite{2604.25884}.
For agents operating on hardware, measured performance provides direct feedback and verification, although repeated trials consume device time.
These agent workflows can use open-source control and simulation libraries to optimize pulses and evaluate proposed control settings~\cite{Lambert2026,Puzzuoli2023,Wittler2021,Goerz2019}.

\textbf{Compiling and optimizing.} Reinforcement learning has been applied to qubit routing and circuit optimization~\cite{2007.15957,2103.07585,2605.10910}, and IBM distributes learned synthesis and routing methods as an add-on to the \texttt{Qiskit} transpiler~\cite{2405.13196}.
Learned search has also reduced CNOT counts~\cite{2604.13812} and T counts~\cite{2402.14396}.
More recently, LLMs have been adapted to synthesize 
Clifford+T circuits 
from unitary operators with gate constraints specified in natural language~\cite{2606.13811}.
Other efforts use LLMs to generate and refine trapped-ion shuttling compilers~\cite{2607.24714} and apply LLM-guided evolutionary search to compiler routines and runtime policies~\cite{2609.38883}.
Efficient classical verification also enables LLM-guided search for quantum circuits with lower Toffoli and qubit counts~\cite{2610.22367,2610.23832}.
The Stabilizer Rank Challenge supports agent-assisted searches for magic-state decompositions verified by exact arithmetic~\cite{StabilizerRankChallenge2026}.
For transformations of existing circuits, equivalence checking can verify that resource reductions preserve the intended computation.
Section~\ref{sec:build} discusses the scope and practical limits of these verification approaches.

\textbf{Discovering error-correcting codes.} Reinforcement learning has been used to discover quantum error-correcting codes and their encoding circuits for specified hardware constraints and noise models~\cite{2305.06378,Olle2024}, and to reduce stabilizer weights~\cite{2502.14372}.
More recently, LLM-based workflows have been used to discover code constructions and assess their correctness and error-correction performance~\cite{2606.02418,2606.24808,2510.20728,2607.25865}.
Inspired by ECDSA.fail~\cite{ECDSAFail2026}, the QEC Challenge supports agent-assisted code discovery with automated verification~\cite{UnitaryAgenticLeaderboards2026,QECChallenge2026}.
Its verification pipeline checks the code structure and searches for low-weight logical operators to place upper bounds on the code distance.
Establishing the exact distance and evaluating logical error rates require further computation. We return to the QEC Challenge in Section~\ref{sec:build} to discuss its role in AI-driven research.

\textbf{Decoding errors.} AI contributes to error decoding through both learned decoders and tools for developing their algorithms and implementations.
Early neural-network decoders~\cite{1610.04238,1705.00857} have been followed by AlphaQubit, a recurrent transformer that reduces logical error rates relative to matching-based decoders on Sycamore surface-code data~\cite{Bausch2024}.
LLM-guided search has also been explored for tuning matching-decoder edge weights~\cite{2604.05587} and optimizing belief-propagation decoder implementations using high-level synthesis feedback~\cite{Bhandwaldar2026}.
In these workflows, LLMs assist decoder development while the resulting algorithms process syndromes during execution.
During development, open-source simulation and decoding packages allow candidate decoders to be compared by their logical error rates under specified noise models without requiring quantum hardware~\cite{2103.02202,Higgott2025,Wu2023}.
Practical deployment requires sufficient decoding throughput and low-latency integration with quantum control hardware~\cite{Bausch2024,NVQLink}.

\textbf{Circuit design and experiments.} LLM-based  workflows have been used to propose and refine variational circuit architectures using simulator feedback~\cite{2307.08191,2602.19387,2606.13380}.
Hardware-aware designs, such as LLM-generated ansätze for quantum circuit Born machines~\cite{2509.08385}, have also been evaluated on quantum processors.
LLMs have further generated Python programs that describe families of photonic experiments and quantum circuits~\cite{Arlt2026}.
Agents also translate scientific objectives into executable experiments and analyze the resulting measurements across several quantum hardware platforms~\cite{2603.08801,2607.25834,2607.25145}.
Simulation-guided design remains limited by the circuit sizes that can be evaluated and by the assumptions of the simulator~\cite{Suzuki2021,Isakov2021}.
For hardware experiments, successful execution must be distinguished from scientifically valid interpretation of the results~\cite{2607.25834,2607.25145}.
Independent evaluation of physical performance is therefore essential~\cite{2609.17439}, as discussed in Section~\ref{sec:build}.

\textbf{Maintenance and debugging.} Maintenance is one of the newest of these functions and the closest in shape to ordinary software work. Migration of code across breaking framework releases has been evaluated for \texttt{Qiskit}~\cite{2606.20173}, and it is the task to which the analysis of Section~\ref{sec:data} applies most directly. AI-assisted maintenance also includes debugging quantum programs. For example, QBugLM evaluates agents on detecting and repairing quantum software bugs~\cite{2606.07314}, and KQFuzz uses codebase knowledge to guide LLM-based fuzzing of quantum libraries~\cite{Xia2026KQFuzz}. These efforts build on earlier studies that cataloged real bugs in quantum frameworks and programs~\cite{Paltenghi2022,Zhao2021} and developed automated detection tools using differential testing, metamorphic testing, and static analysis~\cite{Wang2021,Paltenghi2023,Paltenghi2024}. Beyond these technical evaluations, maintenance also involves submitting code changes as pull requests for human review. This process gives the QOSS community direct experience with AI-assisted contributions and the work required to assess them. We examine this community evidence in Section~\ref{sec:community}.

\section{Constraints in quantum open-source software}
\label{sec:constraints}

QOSS projects share concerns about credit, licensing, training, and access costs with other research software communities~\cite{ReSA2026}. Review burden is a further shared concern, and the clearest recent example comes from \texttt{curl}, which in January 2026 ended its monetary bug bounty after the proportion of valid vulnerability reports fell from above 15\% to below 5\%. Since 2019, the program had paid over US\$100,000 for 87 confirmed vulnerabilities, and its maintainer attributed the decline to AI-generated reports and lower-quality human submissions~\cite{Stenberg2026}. We return to this point in Section~\ref{sec:standards} when discussing contribution requirements.

Beyond these shared concerns, QOSS projects can face additional constraints from limited public code and changing software interfaces. Coding models have fewer framework-specific examples to learn from, and generated code may depend on interfaces that are no longer available. Quantum software also requires verification of physical correctness at scales beyond the reach of exact classical simulation. This verification must establish, for example, that compiler transformations preserve the intended computation and that noise models are implemented correctly. Agents with access to project systems or paid quantum hardware introduce further security and cost risks. 

Our analysis also reveals substantial variation between projects, since quantum and other scientific Python projects overlap in interface stability while quantum frameworks differ in their test sizes and use of randomly generated circuits. These differences limit the usefulness of uniform recommendations for QOSS. Sections~\ref{sec:data}--\ref{sec:pipeline} therefore examine these constraints individually, each concluding with a \textbf{Practice} paragraph offering recommendations that projects can adapt to their own needs. Table~\ref{tab:summary} summarizes the metrics used in this analysis and the conclusions each can support, along with their limitations.

\begin{table*}[t]
\caption{Summary of results and limitations in Sections~\ref{sec:constraints} and \ref{sec:community}. 
Repository data were collected on 6 October 2026. Commit and pull request disclosure rates cover January--September 2026. unitaryHACK records cover the 2022--2026 events. Results from the 2026 QOSS survey use a snapshot collected on 26 October 2026. Methods and project-level results are available in the companion repository~\cite{ai4qoss}. PR denotes pull request. }
\label{tab:summary}
\begin{ruledtabular}
\begin{tabular}{@{}p{\dimexpr0.34\textwidth-2.04\tabcolsep\relax}p{\dimexpr0.13\textwidth-0.78\tabcolsep\relax}p{\dimexpr0.13\textwidth-0.78\tabcolsep\relax}p{\dimexpr0.40\textwidth-2.40\tabcolsep\relax}@{}}
\raggedright Metric & \multicolumn{2}{c}{Result} & \raggedright Notes \tabularnewline
\cline{2-3}
& \raggedright Quantum & \raggedright Non-Quantum & \tabularnewline \colrule
\multicolumn{4}{@{}l}{\textit{Code availability and API stability (Fig.~\ref{fig:corpus})}} \\
\raggedright Public files importing a library & \raggedright 6,864 & \raggedright 1,609,728 & \raggedright 9 quantum and 8 comparison libraries; medians \tabularnewline
\raggedright Monthly PyPI downloads, median & \raggedright 97,269 & \raggedright 191,385,592 & \raggedright Includes repeated downloads from continuous integration \tabularnewline
\raggedright 2021 API still importable & \raggedright \mbox{34.4--71.7\%} & \raggedright \mbox{46.4--85.7\%} & \raggedright Overlapping ranges between groups \tabularnewline
\raggedright Deprecation notices per year & \raggedright 24.1--56.8 & \raggedright 6.8--11.0 & \raggedright Four projects; collected in August 2026\tabularnewline
\colrule
\multicolumn{4}{@{}l}{\textit{Verification and disclosure (Fig.~\ref{fig:verif})}} \\
\raggedright Test qubit counts $\leq 20$ & \raggedright 98.9\% & \raggedright -- & \raggedright Three quantum frameworks; median 2 qubits \tabularnewline
\raggedright Commit AI disclosure, median & \raggedright 0.57\% & \raggedright 1.54\% & \raggedright Attribution lines only; prose disclosures excluded \tabularnewline
\raggedright PR AI disclosure, median & \raggedright 0.43\% & \raggedright 4.56\% & \raggedright Larger variation within groups than between \tabularnewline
\raggedright PR templates requesting AI disclosure & \raggedright 3 of 11 & \raggedright 8 of 8 & \raggedright \tabularnewline
\raggedright Largest contributor share & \raggedright up to 75\% & \raggedright up to 97\% & \raggedright Among projects with at least 5 disclosed commits \tabularnewline
\colrule
\multicolumn{4}{@{}l}{\textit{Candidate tasks and community surveys (Fig.~\ref{fig:tasks})}} \\
\raggedright Candidate maintenance tasks &
\multicolumn{2}{p{\dimexpr0.26\textwidth+0.44\tabcolsep\relax}}{\raggedright 417 of 812} &
\raggedright Distinct issues linked to selected merged PRs \tabularnewline

\raggedright Tasks with test-file changes &
\multicolumn{2}{p{\dimexpr0.26\textwidth+0.44\tabcolsep\relax}}{\raggedright 275 of 417} &
\raggedright Paths identified as tests within the first 100 listed files of each reference PR \tabularnewline

\raggedright Repositories represented &
\multicolumn{2}{p{\dimexpr0.26\textwidth+0.44\tabcolsep\relax}}{\raggedright 110} &
\raggedright Largest repository: 5.0\% of tasks; top five: 17.7\% \tabularnewline
\raggedright Disclosure in free-text responses & \multicolumn{2}{p{\dimexpr0.26\textwidth+0.44\tabcolsep\relax}}{\raggedright 1 of 9 maintainers;\newline 1 of 16 contributors} & \raggedright unitaryHACK responses grouped by topic \tabularnewline
\raggedright AI use for quantum tasks & \multicolumn{2}{p{\dimexpr0.26\textwidth+0.44\tabcolsep\relax}}{\raggedright 715 of 917} & \raggedright Self-reported \tabularnewline
\raggedright Verifying AI-generated code & \multicolumn{2}{p{\dimexpr0.26\textwidth+0.44\tabcolsep\relax}}{\raggedright 79\% manual review;\newline 60\% automated test/verification} & \raggedright 621 respondents; multiple responses allowed \tabularnewline
\end{tabular}
\end{ruledtabular}
\end{table*}

\subsection{Public code availability and interface stability}
\label{sec:data}

\begin{figure*}[t]
  \centering
  \includegraphics[width=\textwidth]{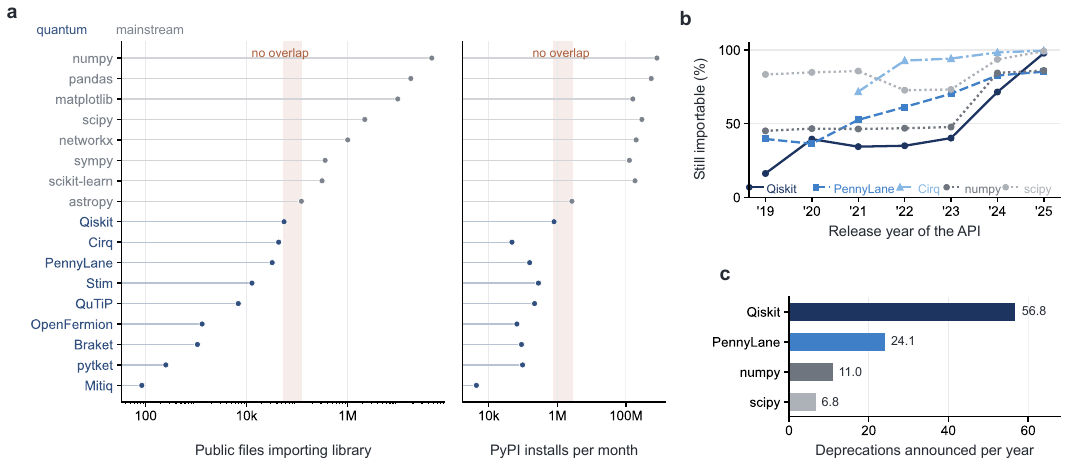}
  \caption{Public code availability and API stability. \textbf{a.} Files in public non-fork repositories that import each library (left) and installs recorded by PyPI over thirty days (right), for nine quantum libraries in blue and eight established scientific Python libraries in gray. The shaded band marks the interval between the two groups, which no library falls into on either measure. \textbf{b.} Fraction of a project's public API of a given year that still imports in its latest 2026 release, counting a symbol as the import path a user would write. \textbf{c.} Deprecations announced per year for the four projects whose release notes carry a deprecation section, taken from those notes. Deprecation counts depend on both API changes and how projects document them. GitHub provides estimated file counts for large search results, while PyPI records package downloads. For panel \textbf{b}, we exclude years in which more than five percent of source files could not be parsed. This excludes \texttt{NumPy} data before 2014 and \texttt{SciPy} data before 2013. Data for panels a and b (c) were collected in October (August) 2026.}
  \label{fig:corpus}
\end{figure*}

We compare quantum frameworks with other scientific Python libraries using counts of public repository files that import each library and package downloads recorded by PyPI. Quantum frameworks have substantially less public code and fewer downloads, with median differences of factors of 235 and 1,968, respectively (Fig.~\ref{fig:corpus}a and Table~\ref{tab:summary}). These measurements characterize differences in ecosystem scale.

By contrast, we find substantial overlap in interface stability between the two groups. Across the projects surveyed, 34 to 86 percent of the 2021 public interface remains importable in the current release (Fig.~\ref{fig:corpus}b). \texttt{Cirq} retains more of its interface than \texttt{NumPy}, while \texttt{SciPy} retains the largest share. We also count deprecations announced in release notes: \texttt{Qiskit} and \texttt{PennyLane} report 56.8 and 24.1 per year, compared with 11.0 for \texttt{NumPy} and 6.8 for \texttt{SciPy} (Fig.~\ref{fig:corpus}c). These counts reflect documentation practices as well as API changes. Compatibility risks therefore need to be assessed for individual frameworks and versions.
As an example, in \texttt{metriq-gym}~\cite{2603.08680}, an older Braket SDK dropped circuit barriers during conversion from Qiskit, inflating benchmark scores and requiring a stricter minimum dependency version~\cite{MetriqGymBarrierFix2026}.

Coding agents can also retrieve documentation, inspect source code, and revise their output using execution feedback, reducing their dependence on knowledge acquired during training. Version-specific retrieval has also been used to assist Qiskit code migration~\cite{2606.20173}. The relevance of public code availability and interface changes to agent performance therefore depends on access to current project information and opportunities to check the generated code.

\textbf{Practice.} As in general open-source development, pin dependency versions and provide coding agents with matching documentation and runnable examples. Verify migrated code using the project's test suite in the target environment~\cite{2606.20173}. To guard against silent physical or semantic regressions across the quantum stack, write explicit dependency-contract tests that report failures explicitly after an upgrade whenever external assumptions change, for example, asserting that cross-SDK circuit conversions preserve non-unitary instructions such as barriers for benchmarking and error mitigation~\cite{MetriqGymBarrierFix2026}, verifying qubit ordering and relative phases on asymmetric entangled states, and checking expected noise-channel fidelities.

\subsection{Verification confined to small circuits}
\label{sec:verif}

Circuit simulation provides reference results for verifying the correctness of quantum programs. However, the memory required for exact state-vector simulation grows exponentially with circuit width, limiting its use for general circuits at large scales~\cite{Suzuki2021,Isakov2021}. Specialized simulators can handle larger structured circuits, but their implementations still require testing at the scales they support. For example, a bug in the stabilizer and matrix product state methods of \texttt{Qiskit Aer} produced correct measurement samples at 64 qubits but incorrect samples at 65 qubits without raising an error~\cite{Aer2107}. Such failures can remain undetected by tests confined to smaller circuits.

\begin{figure*}[t]
  \centering
  \includegraphics[width=\textwidth]{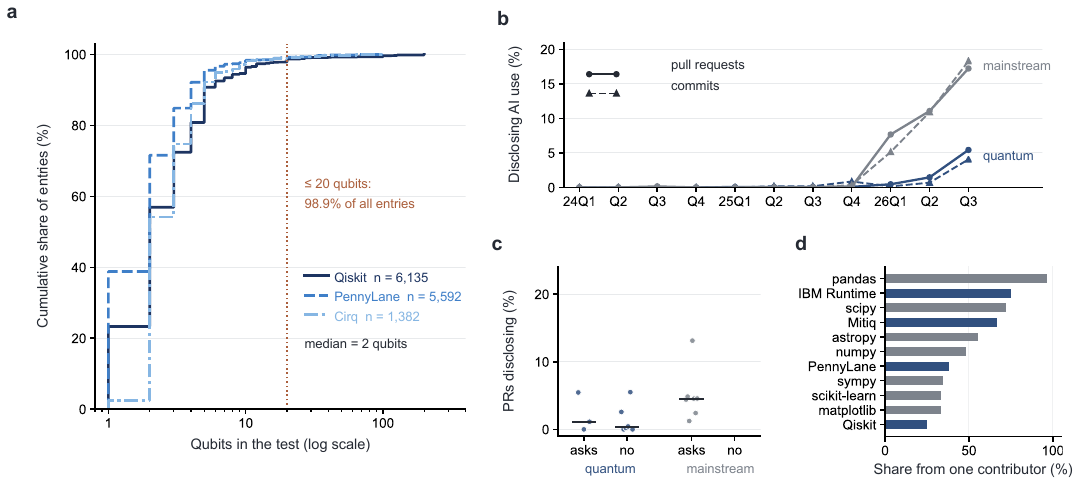}
  \caption{Test circuit sizes and AI disclosure. \textbf{a.} Cumulative distribution of qubit counts extracted from the test code of three quantum frameworks. Here $n$ denotes the number of extracted qubit-count entries, with 13,109 entries in total. \textbf{b.} Share of commits (dashed) and pull requests (solid) carrying an AI disclosure, pooled by quarter within each group, over 11 quantum and 8 mainstream projects. \textbf{c.} Per-project disclosure rate in 2026 pull requests, split by whether the project's pull request template asks contributors to declare AI use; points are projects with at least 100 pull requests that year, lines are group medians, and all eight mainstream projects ask. \textbf{d.} Fraction of a project's disclosing commits written by its single largest discloser, calculated over its available commit history through 30 September 2026. Panel \textbf{a} includes qubit counts extracted from integer values written directly in test code, retaining counts from 1 to 200. Circuit sizes defined through variables or expressions are not evaluated. The disclosure analysis captures AI use recorded in commits or pull requests and misses unreported use. Before identifying disclosures, we remove repeated template text to avoid counting requests to declare AI use as actual disclosures. Test suites are those of October 2026, and disclosure is counted through 30 September 2026.}
  \label{fig:verif}
\end{figure*}

To examine the circuit sizes used in repository tests, we extracted the qubit counts written as literal integers (from 1 to 200) in qubit and wire constructors across the test suites of \texttt{Qiskit}, \texttt{PennyLane}, and \texttt{Cirq}, excluding sizes defined through variables, expressions, or parameterized fixtures. Among the 13,109 extracted qubit-count entries, the median is two qubits, and 98.9 percent are twenty or fewer (Fig.~\ref{fig:verif}a and Table~\ref{tab:summary}). For human-authored unit tests, tests of small circuits are appropriate. Whether a function rejects invalid input, whether a gate has the right matrix, or whether a compiler pass rewrites a local pattern are readily verified with one to three qubits without slowing continuous integration. The question for agentic development is whether repository test suites also include verification at larger circuit sizes in the modules whose behavior depends on scale. Grouped by software function, we find that the 99th-percentile circuit size reaches 120 qubits for simulation and quantum-state operations and 41 for algorithms and applications, compared with 11 for circuit construction and 10 for data import and export. However, the median remains two or three qubits across all categories, and 96.9\% of qubit-count entries from simulation tests are 20 or fewer.

Because we classify test files by path, 3,526 of the 13,109 entries fall outside the named categories, and only five entries fall under error correction, reflecting that large-scale stabilizer and quantum error-correction (QEC) testing is often done against specialized packages such as \texttt{Stim}~\cite{2103.02202} rather than general SDK unit tests. Such tests are often constructed programmatically rather than via literal qubit constructors and are not covered by our methodology. We therefore use the upper quantiles to characterize where larger tests occur rather than to estimate total coverage.

More broadly, small tests alone may not capture errors in methods whose behavior changes with circuit size. Non-local methods, such as block-partitioned synthesis in \texttt{BQSKit}~\cite{BQSKit}, activate partitioning logic only above threshold sizes, accumulate numerical error with scale, and require verification at target widths.

Scaling up verification also requires matching the solver to the circuit structure. When the \texttt{ucc} compiler benchmarked decision-diagram (\texttt{QCEC}~\cite{2004.08420}), ZX-calculus (\texttt{PyZX}~\cite{1904.04735}), and certified-compilation (\texttt{Giallar}~\cite{Tao2022}) verifiers on unstructured random Clifford circuits, none completed 50-qubit instances in the configurations tried, and a 100-qubit \texttt{QCEC} run that exceeded one hour on a 96-core machine was projected to take well over 24 hours~\cite{ucc62}. Yet diagram and ZX verifiers scale well on structurally close compilation pairs, while unitary Clifford circuits admit efficient equivalence
verification through tableau comparison~\cite{2103.02202}, showing that no single verifier replaces state-vector verification across all circuit families.
Test suites also diverge in circuit construction across domains: random generators account for 75\% of matched calls in \texttt{Cirq} and 57\% in \texttt{Qiskit} (suited for fuzzing transpiler passes), versus 5\% in \texttt{PennyLane} (which tests analytical gradients on named states like \texttt{GHZ}, \texttt{Bell}, and \texttt{W} or structured ans\"atze). A problem can arise when coding agents transfer compiler-style random fuzzing to variational algorithms, Hamiltonian simulation, or QEC decoding, where correctness depends on symmetries or syndrome invariants that random circuits never probe.

Furthermore, passing existing unit tests provides evidence for expected behavior on small circuits, but is insufficient to verify behavior across machine-word boundaries, routing on larger topologies, or QEC logical error rates. Yet the practical impact of these verification gaps depends on developer workflows; in Section~\ref{sec:community}, we examine how respondents validate AI-generated code and find they rely on automated tests less often than manual review.

\textbf{Practice.} Use small unit tests to check local correctness, but do not treat them as a sufficient acceptance criterion for coding agents. Supplement them with targeted verification above machine-word and partitioning thresholds (for example, $>64$ qubits), and require agents to pair random-circuit fuzzing for compiler passes with invariant-preserving structured tests for algorithmic and QEC code.

\subsection{Risks of agentic workflows}
\label{sec:agentic}

An agent that opens a pull request in a quantum repository can install packages, edit the continuous-integration configuration, and use credentials for hardware backends where each run costs money. Most of these risks are shared with other automated contributors that have write access. Installing packages can introduce malicious dependencies from public registries~\cite{2005.09535}, while instructions embedded in issues or web pages can redirect an agent’s actions~\cite{2302.12173}. Established security practices for automated development therefore also apply to quantum repositories. Access to paid hardware introduces a financial risk, because repeated retries consume device time and incur charges even when the task remains unresolved. This risk also applies to agents using other paid computing services.

Furthermore, agents operating quantum experiments may also make scientifically consequential choices that successful code execution does not validate. In a neutral-atom workflow, an agent selected an unsuitable observable and produced an incorrect hardware diagnosis, both of which required domain-expert review to identify~\cite{2607.25834}.

\textbf{Practice.} Run agents in isolated environments with task-specific permissions. Require maintainer approval before installing new dependencies or changing verification workflows. Enforce hardware access, spending, and retry limits outside the agent's control. We provide templates for these boundaries in the companion repository~\cite{ai4qoss}.

\subsection{The contributor pipeline}
\label{sec:pipeline}

It has been reported that, in 81.7 percent of the public quantum repositories surveyed, maintenance rests on a single contributor~\cite{Lara2026}. These projects recruit and train future maintainers through good first issues and hackathons. Coding agents could take over this introductory work, reducing opportunities for newcomers to gain experience. Organizers of community hackathon events such as unitaryHACK also report that unverified AI-generated contributions shift maintainer time from mentoring to reviewing submissions~\cite{unitaryHACK2026}. 
Assessing how much of this work agents could take over requires understanding the tasks involved. The 417 unitaryHACK tasks analyzed in Section~\ref{sec:community} include feature development, compatibility work, documentation, and bug fixes. The candidate task collection described in Section~\ref{sec:bench} provides material for future evaluations of these activities.

Completing a contribution with AI assistance also raises the question of what the contributor learns. A randomized study of developers learning an unfamiliar Python library found lower subsequent comprehension and debugging scores with AI assistance~\cite{2601.20245}. For newcomers, pressure to complete tasks quickly may conflict with the time needed to develop these skills~\cite{2607.29178}. Community projects should therefore make room for learning through implementation and debugging, alongside the goal of resolving issues.
The longer-term question is whether participants continue contributing and eventually become maintainers. We present relevant evidence along this direction in Section~\ref{sec:community}.

\textbf{Practice.} Maintainers should configure review agents with project documentation and verification tools, and evaluate them on previously reviewed pull requests. These agents should detect bugs and provide verification results for maintainer review. 
For introductory tasks, contributors should develop an initial approach before requesting generated code and use AI feedback to revise their own work. Maintainers and event organizers should allocate time for mentored implementation and debugging, and recognize learning alongside completed contributions. Follow-up tasks completed without AI assistance could assess whether participants can apply what they learned.

\section{Evidence from the QOSS community}
\label{sec:community}

\subsection{AI disclosure in public repositories}

We first examine the community questions raised in Section~\ref{sec:constraints} using public repository records from 11 quantum projects and 8 other scientific Python projects. We measured how often contributors disclose AI involvement by scanning the attribution lines of commit messages and the text of pull requests.
Specifically,  we compare quarterly disclosure rates for commits and pull requests separately in Fig.~\ref{fig:verif}b. Disclosure is rare in both groups, with the median quantum project carrying a disclosure on 0.43 percent of its 2026 pull requests and the median mainstream project on 4.56 percent (Table~\ref{tab:summary}). We also compare project-level rates according to whether the pull request template asks contributors to disclose AI use, as shown in Fig.~\ref{fig:verif}c. All eight mainstream projects request disclosure, compared with three of the eleven quantum projects, so we interpret recorded disclosure in the context of project policy and as a lower bound on actual use. Individual contributors can also dominate these rates. In one project, 50 percent of 2026 commits carry a disclosure, and 1,107 of its 1,145 disclosing commits come from one maintainer (Fig.~\ref{fig:verif}d). This concentration limits what pooled rates can tell us about typical contributor behavior.

\subsection{Maintenance tasks from community events}

To examine maintainer workload and contributor experiences, we turn to records from unitaryHACK, an annual hackathon that we organize for projects across the quantum open-source community~\cite{unitaryHACK}. The event brings contributors and maintainers together to resolve project issues, develop skills, and introduce newcomers to QOSS. For this study, we compiled annual bounty lists and post-event surveys of contributors and maintainers from 2022 to 2026, which span the years before language models were in general use, the years after ChatGPT, and the current period of coding agents. Across these events, 7 to 14 maintainers and 16 to 33 contributors responded to each annual survey. Our dataset also includes written records of the 2026 maintainer meeting, attended by 27 maintainers. These records provide the basis for our analysis of maintainer workload, contributor experiences, and community responses to AI-assisted development.

The unitaryHACK records also document the maintenance work undertaken by participating projects. Across the five events, we identified 417 distinct candidate tasks from 812 bounty listings, spanning 110 QOSS repositories. Each candidate links an issue description (which might serve as an agent prompt) to a selected merged pull request (which might serve as the reference solution). For 275 candidates, the reference pull request modifies test-related files (Fig.~\ref{fig:tasks}a,b and Table~\ref{tab:summary}). The ratio of retained candidate tasks to bounty listings is 42 percent for 2022 and 65 percent for 2026 (Fig.~\ref{fig:tasks}b).

Feature development and support for software use account for much of this work. Feature and method extensions constitute 225 tasks (54.0\%), while documentation, examples, and presentation-oriented interfaces account for 74 (17.7\%). The remaining tasks include bug fixes, compatibility and integration work, testing, refactoring, performance improvements, and build infrastructure. This composition reflects the work required both to extend software capabilities and to support their use and maintenance. We discuss the collection's technical coverage, its relationship to existing evaluations, and the verification requirements for future agent benchmarking in Section~\ref{sec:bench}.

\subsection{Community experiences with AI-assisted development}

\begin{figure*}[t]
  \centering
  \includegraphics[width=\textwidth]{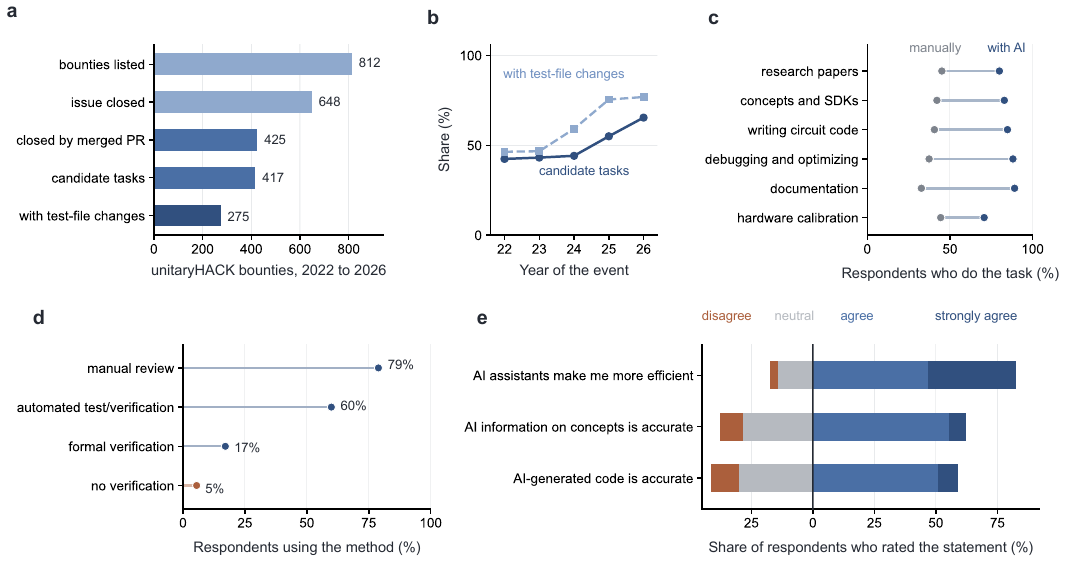}
  \caption{Repository tasks and community use of AI.
  \textbf{a.} Candidate-task extraction from unitaryHACK bounty records spanning 2022--2026. Candidate tasks pair an issue description of at least 120 characters with a merged reference solution. The final bar counts reference solutions containing new or updated test files, identified by path.
  \textbf{b.} Fraction of each year's bounties yielding candidate tasks (solid), and fraction of those tasks whose reference solutions contain tests (dashed).
  \textbf{c.} Reported use of AI assistance and manual work for six quantum workflow tasks. Line length shows the difference between the two. Percentages exclude respondents who marked a task as not applicable. From top to bottom, the denominators are 581, 589, 564, 553, 539 and 232. Respondents could select both modes.
  \textbf{d.} Reported methods for verifying AI-generated code among 621 respondents. Multiple methods could be selected. 
  \textbf{e.} Agreement with statements about the benefits and accuracy of AI assistance. From top to bottom, the denominators are 601, 601, and 597, excluding responses marked not applicable. The two disagreement categories are combined.
  Panels \textbf{c}--\textbf{e} use the 26 October 2026 snapshot of the QOSS survey~\cite{QOSSsurvey2026}, collected while the survey remained open. Percentages describe respondents to each question. Bounty records were resolved against GitHub on 6 October 2026.}
  \label{fig:tasks}
\end{figure*}

To examine AI use beyond individual community events, we analyze responses to the annual Quantum Open Source Software Survey conducted by Unitary Foundation~\cite{QOSSsurvey2026}. We use a snapshot collected on 26 October 2026 comprising 1,691 respondents, while the survey remained open. Among the 917 respondents who answered the question on AI use, 715 reported using AI tools for quantum tasks. Across the six workflow tasks surveyed, 71 to 89 percent of respondents performing each task reported AI assistance (Fig.~\ref{fig:tasks}c). These tasks range from reading research papers and learning quantum concepts to developing code and calibrating hardware. Respondents could report both manual work and AI assistance for a task. These responses describe individual tool use, complementing the contribution-level disclosure records analyzed above.

We also examine how respondents verify AI-generated code and assess its usefulness. Among the 621 respondents to the code-verification question, 79 percent reported manual review, 60 percent automated test or verification, and 17 percent formal verification, while 5 percent reported no verification (Fig.~\ref{fig:tasks}d). Manual review was the most commonly reported method, and respondents could select multiple methods. Agreement was higher for the benefits of AI assistance than for the accuracy of its output. Among those who rated each statement, 83 percent agreed that AI assistants improved the efficiency and effectiveness of their quantum computing work ($n=601$), compared with 59 percent who agreed that generated code was accurate ($n=597$) (Fig.~\ref{fig:tasks}e). These responses support making verification a routine part of AI-assisted development.

To examine the implications for contribution review, we return to the unitaryHACK surveys. In 2026, six of the 10 responding maintainers reported using AI tools in their work, and five reported having policies for AI contributions. We grouped free-text responses from nine maintainers and 16 contributors by topic. Maintainers most often raised human oversight (4 of 9) and review burden (3 of 9). Among contributors, six mentioned review burden and six mentioned contributors' understanding of their changes, while 5 reported that AI tools had helped them. Only one respondent in each group raised disclosure.

When asked about workload after the 2026 unitaryHACK event, 5 of 10 maintainers reported more work than they had anticipated. The corresponding counts were 0 of 8 in 2023, 2 of 7 in 2024, and 4 of 14 in 2025. The free-text responses describe time spent explaining rejections of poor submissions and establishing how AI was used, even in otherwise acceptable contributions. One maintainer described maintaining as talking to an AI tool through the GitHub interface. At the 2026 meeting, maintainers discussed design conversations before pull requests, freezing issues that attracted competing submissions, and office hours before review. The organizers also limited each participant to four open pull requests at a time in 2026 to reduce AI slop.

\section{What the community can build}
\label{sec:build}

The evidence in Sections~\ref{sec:constraints} and \ref{sec:community} identifies practical needs in verification, agent evaluation, and contribution review (Table~\ref{tab:summary}). These needs also arise when agents use QOSS to conduct research. Here we propose tools and practices that help projects assess software contributions and build research environments in which agents can explore candidate solutions under explicit scientific criteria.

To support these proposals, we provide resources that projects can adapt to their own workflows in the companion repository~\cite{ai4qoss}. 
These resources include a verifier catalog documenting tool capabilities and limitations, guidance on collecting and preparing candidate maintenance tasks for future benchmarking, and contribution templates for reporting changes and verification evidence. We also provide draft agent harness designs that describe how agents can use these tools during development and research.
Furthermore, we maintain an annotated reading list to help readers follow the relevant literature as the field develops~\cite{ai4qoss}.

As an illustrative example, our draft harness outlines how an agent could modify a quantum compiler to reduce two-qubit gate counts. The agent would evaluate each proposed change on a fixed circuit set using equivalence checks whose acceptance criteria it cannot modify. Projects can use our verifier catalog to select tools based on their scope and limitations. Contributors can then use our provided templates to report their changes, verification results, and AI assistance for maintainer review.

\subsection{Verification beyond full-state simulation}
\label{sec:verifiers}

To extend verification beyond the small circuits described in Section~\ref{sec:verif}, projects need methods that do not require full-state simulation. Equivalence verifiers such as decision diagrams~\cite{2004.08420} and ZX-calculus rewriting~\cite{1904.04735} can verify whether a compiled circuit preserves the intended computation, including transformations proposed by AI tools. Their worst-case cost remains exponential, so projects should evaluate their runtime on the circuit families and sizes they intend to support. Stabilizer simulation provides another option for Clifford circuits, reaching tens of thousands of qubits~\cite{2103.02202}. This supports verification of large stabilizer error-correction circuits, while non-Clifford operations require other methods such as \texttt{Clifft}~\cite{2604.27058}. Detector error models, which represent errors in quantum error-correction experiments~\cite{2103.02202}, also support efficient equivalence verification. For these models, verification runs in quasilinear time and is exact for non-adaptive pipelines and sound for partially adaptive pipelines such as lattice surgery~\cite{2606.14677}.

Beyond verifying circuits and error models, we can use proof assistants~\cite{Albarghouthi2026Superoptimization,tzap,2605.16523,2607.21533,2510.08672,Peng2023Shor,Hietala2021VOQC,Zhou2023CoqQ,2606.29687} to formalize the mathematical results underlying quantum software. The open-source optimizer \texttt{tzap} includes a Lean 4 implementation with mechanized correctness proofs for core deterministic Clifford+T and $R_z$ optimization passes alongside its Rust implementation~\cite{Albarghouthi2026Superoptimization,tzap}, while also illustrating a practical scope boundary: newer high-level synthesis and heuristic passes in the Rust and Python codebase evolve faster than the formal gate model.
\texttt{Lean-QEC} formalizes stabilizer-code theory in Lean 4 and certifies distance lower bounds for bivariate bicycle codes, including a 90-qubit instance. It uses a verified reduction to Boolean satisfiability and verifies the resulting solver certificates in Lean, removing reliance on unverified solver outputs~\cite{2605.16523}. 
Developing such proofs requires work on formal definitions and supporting lemmas, which can then be reused in later proofs.
Coding agents can help develop these formal libraries by completing proofs. Two Lean 4 benchmarks evaluate this capability through 36 and 40 theorem-completion tasks in quantum algorithms and quantum information theory~\cite{2607.21533}. The best of four models scores 60.4 and 59.6 out of 100, respectively, on a difficulty-weighted scale. Access to a verified domain library improves both scores and completion rates across all eight model--benchmark pairs, with gains of up to 15.9 points. These results support community investment in verified libraries that agents can reuse and extend.

Before adding a verifier to continuous integration, projects should state which properties and circuit families it covers and the sizes at which it remains practical. We maintain a catalog recording this scope, computational cost, and tool maturity in the companion repository~\cite{ai4qoss}. The catalog also identifies priorities for further work. Equivalence verification for non-Clifford circuits beyond exact simulation needs clear soundness guarantees and evidence of practical scaling. Other priorities include distance certification for broader qLDPC code families and automated verification that noise-model parameters satisfy physical constraints. Transpiler passes could also generate correctness certificates for a separate, simpler verifier, removing the transpiler implementation from the components whose correctness must be trusted. Together, these developments would automate more of the verification work currently required of reviewers.

\subsection{Open-source harnesses for AI-driven research}
\label{sec:autoresearch}

A research harness connects an agent to the software that generates, evaluates, and records candidate solutions. These harnesses are being deployed effectively across tasks in quantum information and computing research~\cite{2606.02418,2610.23832,ECDSAFail2026,2609.17439}. We recommend making the scientific objective explicit before choosing a score, since an easily measured improvement may capture only part of the intended research goal. For quantum error correction, for example, code parameters and logical error rates under a specified noise model address different questions. A harness should state which of these it evaluates and the assumptions under which comparisons are meaningful.

As an example, the QEC Challenge illustrates how an open repository can support this process through shared submission rules and automated checks~\cite{QECChallenge2026}. Its verification pipeline checks code structure and searches for logical operators that refute a claimed distance. This supports iterative discovery while leaving a clear distinction between a candidate that survives the search and a code whose distance has been certified. We recommend preserving this distinction in the feedback given to agents and in the results reported to the community.

We further propose separating the agent's search process from the final evaluation, with acceptance criteria that the agent cannot modify~\cite{2609.17439}. Promising candidates should undergo verification appropriate to the claim, with additional tests or stronger certification where needed. We also encourage projects to retain enough information to reproduce each result's construction and evaluation, including the software environment and computational budget. Open-source communities can then contribute improved evaluators and checks alongside new candidates, allowing the research process itself to be inspected and refined.

\subsection{Benchmarks built from real repository work}
\label{sec:bench}

QOSS maintenance involves modifying existing repositories beyond short programs and functions. SWE-bench evaluates this type of work by asking agents to resolve GitHub issues and assessing their changes with repository tests~\cite{2310.06770}. AInsteinBench applies repository-level evaluation to scientific software across six codebases, including Qiskit~\cite{2512.21373}. Quantum Arena further provides a public leaderboard comparing general-purpose LLMs and quantum-specific coding assistants on quantum-programming benchmarks~\cite{QuantumArena2026}. The pilot collection described in Section~\ref{sec:community} contains 417 candidate maintenance tasks from 110 QOSS repositories across five unitaryHACK events. Across the candidate tasks, selected reference changes have a median of approximately 300 modified lines. The collection's coverage of feature development, documentation, and other maintenance activities provides material for evaluating how agents extend and maintain existing projects.

The tasks span several parts of the quantum software stack, with uneven representation across domains. Circuit, operator, and intermediate-representation tasks account for 141 cases, and simulation or numerical methods for 86. Error correction and error mitigation account for 12 and 10 tasks, respectively, and domain labels can overlap. The frameworks involved include \texttt{Qiskit}, \texttt{PennyLane}, and \texttt{Cirq}, with \texttt{Qiskit} appearing directly in 53 tasks across 26 repositories. Python files are modified in 76.5\% of tasks, alongside changes in Julia, Rust, C++, and other languages. These file-based counts include test and documentation changes. Simulator involvement is recorded for 96 tasks and provider-interface involvement for 24, including Amazon Braket, IonQ, and IBM Quantum. Provider-interface involvement includes mocked API tests and does not establish execution on quantum hardware. This distribution reflects the participating projects and their bounty choices.

The variety of maintenance activities requires verification matched to each task's requirements. A merged pull request may address only part of an issue, and changes to test files do not establish complete coverage or scientific correctness. For circuit transformations intended to preserve the computation, equivalence verification can assess correctness. Numerical methods require appropriate reference values or invariants, and performance changes require correctness and resource measurements. Documentation and interface tasks may combine executable examples, build or interaction tests, and maintainer review. Future work should establish reproducible project environments and task-specific verification criteria before evaluating agents on the collection.

Once these criteria are established, the tasks could also provide a way to study the role of verification. For circuit-transformation tasks, we can compare an agent given repository tests alone with one given additional equivalence verification during development. Holding the agent configuration, computational budget, and final evaluation criteria fixed would allow us to measure how the additional verification affects the resulting solutions. These comparisons require scoring criteria that reflect the intended capability. For example, Quantum-Harbor verifies scientific results against recorded data or fresh hidden evaluations and distinguishes scientific errors from submission failures~\cite{2609.17439}. Repository benchmarks should likewise accept valid alternative implementations and isolate reference solutions and held-out evaluation data from the agent. Future benchmark evaluations should also examine how programming language affects agent performance. Comparisons on similar tasks with a fixed agent configuration could help distinguish the effects of compiler feedback from differences in model familiarity with each language and its libraries. Such evidence could inform language choices for new QOSS projects alongside runtime performance and maintenance needs.

Annual community events such as unitaryHACK provide a recurring source of candidate tasks for future benchmark updates. We recommend collecting and archiving issue descriptions, merged reference solutions, and associated tests as tasks are completed, preserving these records as repositories change (Fig.~\ref{fig:tasks}b). For each model, evaluations should prioritize tasks resolved after its training cutoff to reduce contamination from previously published solutions~\cite{2506.12286}. Task recency alone does not establish freedom from training contamination. We provide the collection and classification methods and aggregate results in the companion repository~\cite{ai4qoss}, while the task list remains unreleased. The collection method can also be applied to other projects with linked issues and merged pull requests.

\subsection{Standards for AI contributions}
\label{sec:standards}

AI should reduce the cost of contributing to and using open-source software without lowering the standards of responsibility, evidence, and reciprocity that make open source work. This is particularly important in quantum software, where AI-generated claims may rely on subtle mathematical assumptions or physical intuitions, making hallucinations difficult to detect and increasing the cost of verification. Open-source implementations and reproducible results help make these claims independently verifiable, but this openness must be matched by good research and development hygiene: clearly documenting assumptions, rigorously checking results, and giving proper credit to the people whose ideas, methods, and code make the work possible.

Contribution standards should require authors to explain the value of a result and provide evidence that others can assess without repeating the entire investigation.
Our unitaryHACK surveys identify review workload, human oversight, and contributors' understanding of their changes as recurring concerns (Section~\ref{sec:community}). The broader QOSS survey finds that manual review is the most commonly reported method for verifying AI-generated code, with automated tests reported less often. Together, these findings motivate contribution requirements that provide reviewers with reproducible verification evidence. We recommend applying these requirements to all contributions, including those developed with AI assistance.

To put these requirements into practice, we suggest that contributors should explain the intended outcome of a change and provide reproducible verification evidence. Bug reports and fixes should include reproducing examples to reduce the review burden illustrated by the \texttt{curl} case in Section~\ref{sec:constraints}~\cite{Stenberg2026}. New features or methods should include verification against stated functional or performance requirements. Automated workflows should run the required verification procedures before maintainer review. Project-specific review agents can detect bugs and summarize verification results, as discussed in Section~\ref{sec:pipeline}. Maintainers can then assess whether the evidence supports the proposed change and decide whether to accept it.

Alongside verification evidence, contributions should record which AI tools were used and which parts of the work they assisted. Tool names and versions can be recorded automatically where available to support later investigation of tool-related problems. Acceptance should depend on the quality of the contribution and its verification evidence. Contributors remain responsible for explaining their changes and responding to issues raised during review~\cite{VanDerWalt2026AI,Hannon2026AI}. We provide contribution templates and configurations for automated verification in the companion repository~\cite{ai4qoss}.

Finally, we propose taking verified QOSS contributions as a form of research publication on equal footing with journal or conference articles. For certain quantum research results such as computational methods, a versioned software release with reproducible evidence may communicate the result more effectively than a standalone paper. Such contributions should explain their scientific significance and limitations and provide clear attribution, so that researchers and agents can assess and reuse them.

\section{Outlook}
\label{sec:outlook}

Our analysis identifies verification and maintenance as central to the practical use of AI in QOSS.
We draw on project data and community records, and characterize a pilot collection of candidate maintenance tasks for future agent evaluation. The collection identifies maintenance activities and verification requirements that future studies could examine alongside the review effort required from maintainers.
The broader goal is to support both reliable quantum software and AI-driven research whose results the community can independently assess and build upon.

For quantum software development, we propose that future evaluations should establish how model capability, access to project knowledge, and verification affect the quality and cost of agent-generated changes. Comparisons on maintenance tasks with common final evaluation criteria could determine whether additional verification improves correctness and reduces subsequent review effort. Sharing verified examples and formal libraries would allow projects to reuse this work while retaining their own interfaces and development practices. These evaluations should also account for bugs discovered after merging and the effort required to maintain accepted changes.

For AI-driven auto-research, QOSS can provide environments in which agents explore open problems in quantum information and computation~\cite{qiqcop_zoo,2608.28433}. We further propose evaluating whether agents can select productive research directions and revise their approaches in response to unsuccessful attempts. Such evaluations should establish whether the resulting findings remain valid beyond the conditions explored during the search and can be independently reproduced. Recording computational costs alongside scientific outcomes would help identify where agent-driven exploration is useful and which improvements to software and verification tools would most benefit further research.

Sustaining both directions will require attention to the communities that develop and maintain this infrastructure. We recommend assessing changes in review effort alongside opportunities for contributors to learn and take responsibility for their work. Review agents could help maintainers devote more time to design and mentoring. Our October 2026 analyses provide a baseline for tracking maintenance activity through public repository records, complemented by surveys of review effort and contributor experiences. We also propose using these community records and evaluation resources for metascience research on how AI changes scientific practice.   Given the rapid development of the field, we maintain supporting resources in an open companion repository so that the analyses and proposals can evolve with frameworks and AI capabilities~\cite{ai4qoss}. These efforts will help the community develop reliable quantum software and verifiable AI-driven research while sustaining the expertise needed to support both.

\section*{Acknowledgments}
We thank Travis L. Scholten (IBM Quantum and Unitary Foundation Board of Directors) for his valuable feedback on earlier drafts of this manuscript.
We thank Mateo Uldemolins for helpful feedback on the manuscript. M.R.'s work has been supported by the Mozilla Foundation. We used OpenAI Codex and Claude Code to assist with literature searches, manuscript preparation, and the development of data-analysis and plotting scripts.

\section*{Data and code availability}

Data for the figures are available in the companion repository at \url{https://github.com/unitaryfoundation/ai4qoss}. The repository also provides methodological descriptions and example code and configurations for automated verification. For the candidate task collection, we provide collection and classification methods, category definitions, and aggregate statistics.

\makeatletter
\if@filesw
  \immediate\write\@auxout{\string\citation{REVTEX42Control}}
\fi
\makeatother
\bibliographystyle{apsrev4-2}
\bibliography{main}

@CONTROL{REVTEX42Control,
  title = {0},
  eprint = {0},
}

@article{1610.04238,
  author = {Giacomo Torlai and Roger G. Melko},
  title = {{Neural Decoder for Topological Codes}},
  journal = {Phys. Rev. Lett.},
  volume = {119},
  pages = {030501},
  year = {2017},
  doi = {10.1103/PhysRevLett.119.030501},
  number = {3},
}

@article{Miao2026Paper2Agent,
  author = {Jiacheng Miao and Joe R. Davis and Yaohui Zhang and Jonathan K. Pritchard and James Zou},
  title = {Reimagining research papers as interactive and reliable {AI} agents},
  journal = {Nature},
  year = {2026},
  month = {sep},
  doi = {10.1038/s41586-026-11044-y},
  url = {https://www.nature.com/articles/s41586-026-11044-y},
}

@misc{QuantumArena2026,
  author = {{Quantum Science Center}},
  title = {{Quantum Arena}},
  year = {2026},
  howpublished = {\url{https://qscsoftwareecosystem.github.io/quantum-leaderboard/index.html}},
  note = {accessed 22 October 2026},
}

@article{1705.00857,
  author = {Savvas Varsamopoulos and Ben Criger and Koen Bertels},
  title = {{Decoding small surface codes with feedforward neural networks}},
  journal = {Quantum Sci. Technol.},
  volume = {3},
  pages = {015004},
  year = {2018},
  doi = {10.1088/2058-9565/aa955a},
  number = {1},
}

@article{1803.01857,
  author = {Murphy Yuezhen Niu and Sergio Boixo and Vadim N. Smelyanskiy and Hartmut Neven},
  title = {{Universal quantum control through deep reinforcement learning}},
  journal = {npj Quantum Inf.},
  volume = {5},
  pages = {33},
  year = {2019},
  doi = {10.1038/s41534-019-0141-3},
  number = {1},
}

@misc{2605.10910,
  author = {Richie Yeung and Aleks Kissinger and Rob Cornish},
  title = {Equivariant Reinforcement Learning for {Clifford} Quantum Circuit Synthesis},
  year = {2026},
  eprint = {2605.10910},
  archivePrefix = {arXiv},
  primaryClass = {quant-ph},
}

@misc{Albarghouthi2026Superoptimization,
  author = {Aws Albarghouthi},
  title = {Fast Quantum-Circuit Superoptimization},
  howpublished = {Cryptology ePrint Archive, Paper 2026/2115},
  year = {2026},
  url = {https://eprint.iacr.org/2026/2115},
  note = {accessed 22 October 2026},
}

@misc{tzap,
      title={Linear-Time {T}-Gate Optimization via Random Abstraction}, 
      author={Aws Albarghouthi},
      year={2026},
      eprint={2605.13929},
      archivePrefix={arXiv},
      primaryClass={cs.PL},
      url={https://arxiv.org/abs/2605.13929}, 
}

@misc{2601.20245,
  title = {How {AI} Impacts Skill Formation},
  author = {Judy Hanwen Shen and Alex Tamkin},
  year = {2026},
  eprint = {2601.20245},
  archivePrefix = {arXiv},
  primaryClass = {cs.CY},
}

@misc{1803.03226,
  author = {Julian Kelly and Peter O'Malley and Matthew Neeley and Hartmut Neven and John M. Martinis},
  title = {{Physical qubit calibration on a directed acyclic graph}},
  year = {2018},
  eprint = {1803.03226},
  archivePrefix = {arXiv},
  primaryClass = {quant-ph},
}

@misc{1811.04968,
  author = {Ville Bergholm and Josh Izaac and Maria Schuld and Christian Gogolin and Shahnawaz Ahmed and Vishnu Ajith and M. Sohaib Alam and Guillermo Alonso-Linaje and B. AkashNarayanan and Ali Asadi and Juan Miguel Arrazola and Utkarsh Azad and Sam Banning and Carsten Blank and Thomas R Bromley and Benjamin A. Cordier and Jack Ceroni and Alain Delgado and Di Matteo, Olivia and Amintor Dusko and Tanya Garg and Diego Guala and Anthony Hayes and Ryan Hill and Aroosa Ijaz and Theodor Isacsson and David Ittah and Soran Jahangiri and Prateek Jain and Edward Jiang and Ankit Khandelwal and Korbinian Kottmann and Robert A. Lang and Christina Lee and Thomas Loke and Angus Lowe and Keri McKiernan and Johannes Jakob Meyer and J. A. Montañez-Barrera and Romain Moyard and Zeyue Niu and Lee James O'Riordan and Steven Oud and Ashish Panigrahi and Chae-Yeun Park and Daniel Polatajko and Nicolás Quesada and Chase Roberts and Nahum Sá and Isidor Schoch and Borun Shi and Shuli Shu and Sukin Sim and Arshpreet Singh and Ingrid Strandberg and Jay Soni and Antal Száva and Slimane Thabet and Rodrigo A. Vargas-Hernández and Trevor Vincent and Nicola Vitucci and Maurice Weber and David Wierichs and Roeland Wiersema and Moritz Willmann and Vincent Wong and Shaoming Zhang and Nathan Killoran},
  title = {{PennyLane: Automatic differentiation of hybrid quantum-classical computations}},
  year = {2018},
  eprint = {1811.04968},
  archivePrefix = {arXiv},
  primaryClass = {quant-ph},
}

@article{1904.04735,
  author = {Aleks Kissinger and John van de Wetering},
  title = {{PyZX: Large Scale Automated Diagrammatic Reasoning}},
  journal = {Electron. Proc. Theor. Comput. Sci.},
  volume = {318},
  pages = {229--241},
  year = {2020},
  doi = {10.4204/EPTCS.318.14},
}

@article{2004.08420,
  author = {Lukas Burgholzer and Robert Wille},
  title = {{Advanced Equivalence Checking for Quantum Circuits}},
  journal = {IEEE Trans. Comput.-Aided Des. Integr. Circuits Syst.},
  volume = {40},
  pages = {1810--1824},
  year = {2021},
  doi = {10.1109/TCAD.2020.3032630},
  number = {9},
}

@article{2007.15957,
  author = {Matteo G. Pozzi and Steven J. Herbert and Akash Sengupta and Robert D. Mullins},
  title = {{Using Reinforcement Learning to Perform Qubit Routing in Quantum Compilers}},
  journal = {ACM Trans. Quantum Comput.},
  volume = {3},
  pages = {1--25},
  year = {2022},
  doi = {10.1145/3520434},
  number = {2},
}

@article{2103.02202,
  author = {Craig Gidney},
  title = {{Stim: a fast stabilizer circuit simulator}},
  journal = {Quantum},
  volume = {5},
  pages = {497},
  year = {2021},
  doi = {10.22331/q-2021-07-06-497},
}

@misc{2103.07585,
  author = {Thomas Fösel and Murphy Yuezhen Niu and Florian Marquardt and Li Li},
  title = {{Quantum circuit optimization with deep reinforcement learning}},
  year = {2021},
  eprint = {2103.07585},
  archivePrefix = {arXiv},
  primaryClass = {quant-ph},
}

@article{2104.14539,
  author = {V. V. Sivak and A. Eickbusch and H. Liu and B. Royer and I. Tsioutsios and M. H. Devoret},
  title = {{Model-Free Quantum Control with Reinforcement Learning}},
  journal = {Phys. Rev. X},
  volume = {12},
  pages = {011059},
  year = {2022},
  doi = {10.1103/PhysRevX.12.011059},
  number = {1},
}

@article{2105.01079,
  author = {Yuval Baum and Mirko Amico and Sean Howell and Michael Hush and Maggie Liuzzi and Pranav Mundada and Thomas Merkh and Andre R. R. Carvalho and Michael J. Biercuk},
  title = {{Experimental Deep Reinforcement Learning for Error-Robust Gate-Set Design on a Superconducting Quantum Computer}},
  journal = {PRX Quantum},
  volume = {2},
  pages = {040324},
  year = {2021},
  doi = {10.1103/PRXQuantum.2.040324},
  number = {4},
}

@inproceedings{2302.12173,
  author = {Kai Greshake and Sahar Abdelnabi and Shailesh Mishra and Christoph Endres and Thorsten Holz and Mario Fritz},
  title = {{Not What You've Signed Up For: Compromising Real-World LLM-Integrated Applications with Indirect Prompt Injection}},
  year = {2023},
  doi = {10.1145/3605764.3623985},
  booktitle = {Proceedings of the 16th ACM Workshop on Artificial Intelligence and Security},
  pages = {79--90},
  publisher = {ACM},
}

@inproceedings{2310.06770,
  author = {Carlos E. Jimenez and John Yang and Alexander Wettig and Shunyu Yao and Kexin Pei and Ofir Press and Karthik Narasimhan},
  title = {{SWE-bench: Can Language Models Resolve Real-World GitHub Issues?}},
  year = {2024},
  booktitle = {The Twelfth International Conference on Learning Representations},
  url = {https://openreview.net/forum?id=VTF8yNQM66},
}

@article{2402.14396,
  author = {Francisco J. R. Ruiz and Tuomas Laakkonen and Johannes Bausch and Matej Balog and Mohammadamin Barekatain and Francisco J. H. Heras and Alexander Novikov and Nathan Fitzpatrick and Bernardino Romera-Paredes and John van de Wetering and Alhussein Fawzi and Konstantinos Meichanetzidis and Pushmeet Kohli},
  title = {{Quantum circuit optimization with AlphaTensor}},
  journal = {Nat. Mach. Intell.},
  volume = {7},
  pages = {374--385},
  year = {2025},
  doi = {10.1038/s42256-025-01001-1},
  number = {3},
}

@misc{2405.08810,
  author = {Ali Javadi-Abhari and Matthew Treinish and Kevin Krsulich and Christopher J. Wood and Jake Lishman and Julien Gacon and Simon Martiel and Paul D. Nation and Lev S. Bishop and Andrew W. Cross and Blake R. Johnson and Jay M. Gambetta},
  title = {{Quantum computing with Qiskit}},
  year = {2024},
  eprint = {2405.08810},
  archivePrefix = {arXiv},
  primaryClass = {quant-ph},
}

@misc{2405.13196,
  author = {David Kremer and Victor Villar and Hanhee Paik and Ivan Duran and Ismael Faro and Juan Cruz-Benito},
  title = {{Practical and efficient quantum circuit synthesis and transpiling with Reinforcement Learning}},
  year = {2024},
  eprint = {2405.13196},
  archivePrefix = {arXiv},
  primaryClass = {quant-ph},
}

@inproceedings{2405.19495,
  author = {Nicolas Dupuis and Luca Buratti and Sanjay Vishwakarma and Aitana Viudes Forrat and David Kremer and Ismael Faro and Ruchir Puri and Juan Cruz-Benito},
  title = {{Qiskit Code Assistant: Training LLMs for generating Quantum Computing Code}},
  year = {2024},
  doi = {10.1109/lad62341.2024.10691762},
  booktitle = {2024 IEEE LLM Aided Design Workshop (LAD)},
  pages = {1--4},
  publisher = {IEEE},
}

@inproceedings{2406.14712,
  author = {Sanjay Vishwakarma and Francis Harkins and Siddharth Golecha and Vishal Sharathchandra Bajpe and Nicolas Dupuis and Luca Buratti and David Kremer and Ismael Faro and Ruchir Puri and Juan Cruz-Benito},
  title = {{Qiskit HumanEval: An Evaluation Benchmark for Quantum Code Generative Models}},
  booktitle = {{2024 IEEE International Conference on Quantum Computing and Engineering (QCE)}},
  publisher = {IEEE},
  pages = {1169--1176},
  year = {2024},
  doi = {10.1109/QCE60285.2024.00137},
}

@article{2411.09131,
  author = {Yuri Alexeev and Marwa H. Farag and Taylor L. Patti and Mark E. Wolf and Natalia Ares and Alán Aspuru-Guzik and Simon C. Benjamin and Zhenyu Cai and Shuxiang Cao and Christopher Chamberland and Zohim Chandani and Federico Fedele and Ikko Hamamura and Nicholas Harrigan and Jin-Sung Kim and Elica Kyoseva and Justin G. Lietz and Tom Lubowe and Alexander McCaskey and Roger G. Melko and Kouhei Nakaji and Alberto Peruzzo and Pooja Rao and Bruno Schmitt and Sam Stanwyck and Norm M. Tubman and Hanrui Wang and Timothy Costa},
  title = {{Artificial intelligence for quantum computing}},
  year = {2025},
  doi = {10.1038/s41467-025-65836-3},
  journal = {Nature Communications},
  volume = {16},
  number = {1},
  eid = {10829},
}

@article{2412.07978,
  author = {Shuxiang Cao and Zijian Zhang and Mohammed Alghadeer and Simone D Fasciati and Michele Piscitelli and Mustafa Bakr and Peter Leek and Alán Aspuru-Guzik},
  title = {{Automating quantum computing laboratory experiments with an agent-based AI framework}},
  year = {2025},
  doi = {10.1016/j.patter.2025.101372},
  journal = {Patterns},
  volume = {6},
  number = {10},
  eid = {101372},
  pages = {101372},
}

@inproceedings{2503.02497,
  author = {Abdul Basit and Nouhaila Innan and Muhammad Haider Asif and Minghao Shao and Muhammad Kashif and Alberto Marchisio and Muhammad Shafique},
  title = {{A PennyLane-Centric Dataset to Enhance LLM-based Quantum Code Generation using RAG}},
  year = {2026},
  doi = {10.1109/ijcnn69854.2026.11729647},
  booktitle = {2026 International Joint Conference on Neural Networks (IJCNN)},
  pages = {1--8},
  publisher = {IEEE},
}

@inproceedings{2506.12286,
  author = {Shanchao Liang and Spandan Garg and Roshanak {Zilouchian Moghaddam}},
  title = {{The SWE-Bench Illusion: When State-of-the-Art LLMs Remember Instead of Reason}},
  booktitle = {{Proceedings of the IEEE/ACM 48th International Conference on Software Engineering: Software Engineering in Practice (ICSE-SEIP)}},
  publisher = {ACM},
  pages = {395--405},
  year = {2026},
  doi = {10.1145/3786583.3786882},
}

@misc{2510.08672,
  author = {Alex Meiburg and Leonardo A. Lessa and Rodolfo R. Soldati},
  title = {{A Formalization of the Generalized Quantum Stein's Lemma in Lean}},
  year = {2025},
  eprint = {2510.08672},
  archivePrefix = {arXiv},
  primaryClass = {quant-ph},
}

@misc{2511.00092,
  author = {Shunya Minami and Tatsuya Ishigaki and Ikko Hamamura and Taku Mikuriya and Youmi Ma and Naoaki Okazaki and Hiroya Takamura and Yohichi Suzuki and Tadashi Kadowaki},
  title = {{QuantumBench: A Benchmark for Quantum Problem Solving}},
  year = {2025},
  eprint = {2511.00092},
  archivePrefix = {arXiv},
  primaryClass = {cs.AI},
}

@misc{2512.21373,
  author = {Titouan Duston and Shuo Xin and Yang Sun and Daoguang Zan and Aoyan Li and Shulin Xin and Kai Shen and Yixiao Chen and Qiming Sun and Ge Zhang and Jiashuo Liu and Huan Zhou and Jingkai Liu and Zhichen Pu and Yuanheng Wang and Bo-Xuan Ge and Xin Tong and Fei Ye and Zhi-Chao Zhao and Wen-Biao Han and Zhoujian Cao and Yueran Zhao and Weiluo Ren and Qingshen Long and Yuxiao Liu and Anni Huang and Yidi Du and Yuanyuan Rong and Jiahao Peng},
  title = {{AInsteinBench: Benchmarking Coding Agents on Scientific Repositories}},
  year = {2025},
  eprint = {2512.21373},
  archivePrefix = {arXiv},
  primaryClass = {cs.SE},
}

@misc{2602.19387,
  author = {Marco Knipfer and Alexander Roman and Konstantin T. Matchev and Katia Matcheva and Sergei Gleyzer},
  title = {{AI Agents for Variational Quantum Circuit Design}},
  year = {2026},
  eprint = {2602.19387},
  archivePrefix = {arXiv},
  primaryClass = {quant-ph},
}

@misc{2604.08570,
  author = {Ali Slim and Haydar Hamieh and Jawad Kotaich and Yehya Ghosn and Mahdi Chehimi and Ammar Mohanna and Hasan Abed Al Kader Hammoud and Bernard Ghanem},
  title = {{QuanBench+: A Unified Multi-Framework Benchmark for LLM-Based Quantum Code Generation}},
  year = {2026},
  eprint = {2604.08570},
  archivePrefix = {arXiv},
  primaryClass = {cs.LG},
}

@misc{2604.13812,
  author = {Jacopo Cossio and Daniele Lizzio Bosco and Riccardo Romanello and Giuseppe Serra and Carla Piazza},
  title = {{AlphaCNOT: Learning CNOT Minimization with Model-Based Planning}},
  year = {2026},
  eprint = {2604.13812},
  archivePrefix = {arXiv},
  primaryClass = {cs.AI},
}

@misc{2604.27058,
  author = {Bradley A. Chase and Farrokh Labib},
  title = {{Clifft: Fast Exact Simulation of Near-Clifford Quantum Circuits}},
  year = {2026},
  eprint = {2604.27058},
  archivePrefix = {arXiv},
  primaryClass = {quant-ph},
}

@misc{2605.16523,
  author = {Mattias Ehatamm and Yi Lee and Xiaodi Wu and Runzhou Tao},
  title = {{End-to-End Formalization of Quantum Error Correction}},
  year = {2026},
  eprint = {2605.16523},
  archivePrefix = {arXiv},
  primaryClass = {quant-ph},
}

@misc{2605.26021,
  author = {Yusheng Zhao and Han Wang and Xin Liu and Xinjie Song and Jixi He and Lingwei Song and Yuanhe Ji and Ken Deng and Runqing Zhang and Zhiguo Huang and Ling Qian and Jize Han and Di Luo},
  title = {{Toward General Quantum Control with Physics-Informed Large Language Models}},
  year = {2026},
  eprint = {2605.26021},
  archivePrefix = {arXiv},
  primaryClass = {quant-ph},
}

@misc{2605.27210,
  author = {Juan Cruz-Benito and Ismael Faro},
  title = {{Qiskit QuantumKatas: Adapting Microsoft's Quantum Computing exercises for LLM evaluation}},
  year = {2026},
  eprint = {2605.27210},
  archivePrefix = {arXiv},
  primaryClass = {quant-ph},
}

@inproceedings{2606.07314,
  author = {An B. B. Pham and Hoa T. Nguyen and Muhammad Usman},
  title = {{QBugLM: An Agentic Benchmarking Framework for LLM-based Quantum Software Debugging}},
  year = {2026},
  doi = {10.1109/qsw72780.2026.00028},
  booktitle = {2026 IEEE International Conference on Quantum Software (QSW)},
  pages = {185--191},
  publisher = {IEEE},
}

@misc{2606.13380,
  author = {Kenya Sakka and Wataru Mizukami and Kosuke Mitarai},
  title = {{An LLM System for Autonomous Variational Quantum Circuit Design}},
  year = {2026},
  eprint = {2606.13380},
  archivePrefix = {arXiv},
  primaryClass = {quant-ph},
}

@misc{2606.14677,
  author = {Mathys Rennela},
  title = {{Quasilinear Equivalence Checking for Detector Error Models}},
  year = {2026},
  eprint = {2606.14677},
  archivePrefix = {arXiv},
  primaryClass = {quant-ph},
}

@misc{2606.20173,
  author = {Jose Manuel Suarez and Luis Mariano Bibbo and Joaquin Bogado and Alejandro Fernandez},
  title = {{Qiskit Code Migration with LLMs}},
  year = {2026},
  eprint = {2606.20173},
  archivePrefix = {arXiv},
  primaryClass = {cs.SE},
}

@misc{2606.29687,
  author = {Uri Kol and Maor Ben-Shahar and Kfir Sulimany and Dirk Englund},
  title = {{A Machine-Verified Proof of a Quantum-Optimization Conjecture}},
  year = {2026},
  eprint = {2606.29687},
  archivePrefix = {arXiv},
  primaryClass = {quant-ph},
}

@misc{2607.21533,
  author = {Lei Zhang and Yusheng Zhao and Yimeng Cao and Ranyiliu Chen and Mingrui Jing and Jizhe Lai and Ziao Tang and Jingu Xie and Hongshun Yao and Xuanqiang Zhao and Guocheng Zhen and Chengkai Zhu and Xin Wang},
  title = {{Benchmarking Agents for Proving Theorems in Quantum Algorithms and Quantum Information}},
  year = {2026},
  eprint = {2607.21533},
  archivePrefix = {arXiv},
  primaryClass = {quant-ph},
}

@inproceedings{2005.09535,
  author = {Marc Ohm and Henrik Plate and Arnold Sykosch and Michael Meier},
  title = {{Backstabber's Knife Collection: A Review of Open Source Software Supply Chain Attacks}},
  booktitle = {{Detection of Intrusions and Malware, and Vulnerability Assessment (DIMVA 2020)}},
  series = {Lecture Notes in Computer Science},
  publisher = {Springer},
  pages = {23--43},
  year = {2020},
  doi = {10.1007/978-3-030-52683-2_2},
}

@inproceedings{Tao2022,
  author = {Runzhou Tao and Yunong Shi and Jianan Yao and Xupeng Li and Ali Javadi-Abhari and Andrew W. Cross and Frederic T. Chong and Ronghui Gu},
  title = {{Giallar: Push-button verification for the Qiskit quantum compiler}},
  booktitle = {{Proceedings of the 43rd ACM SIGPLAN International Conference on Programming Language Design and Implementation (PLDI 2022)}},
  publisher = {ACM},
  pages = {641--656},
  year = {2022},
  doi = {10.1145/3519939.3523431},
}

@article{Bausch2024,
  author = {Johannes Bausch and Andrew W. Senior and Francisco J. H. Heras and Thomas Edlich and Alex Davies and Michael Newman and Cody Jones and Kevin Satzinger and Murphy Yuezhen Niu and Sam Blackwell and George Holland and Dvir Kafri and Juan Atalaya and Craig Gidney and Demis Hassabis and Sergio Boixo and Hartmut Neven and Pushmeet Kohli},
  title = {{Learning high-accuracy error decoding for quantum processors}},
  journal = {Nature},
  volume = {635},
  pages = {834--840},
  year = {2024},
  doi = {10.1038/s41586-024-08148-8},
  number = {8040},
}

@article{Sivak2026,
  author = {Volodymyr Sivak and Alexis Morvan and Michael Broughton and others},
  title = {{Reinforcement learning control of quantum error correction}},
  journal = {Nature},
  volume = {655},
  pages = {879--884},
  year = {2026},
  doi = {10.1038/s41586-026-10759-2},
}

@article{Pan2025,
	Author = {Pan, Haining and Mudur, Nayantara and Taranto, William and Tikhanovskaya, Maria and Venugopalan, Subhashini and Bahri, Yasaman and Brenner, Michael P. and Kim, Eun-Ah},
	Da = {2025/01/31},
	Doi = {10.1038/s42005-025-01956-y},
	Id = {Pan2025},
	Isbn = {2399-3650},
	Journal = {Communications Physics},
	Number = {1},
	Pages = {49},
	Title = {Quantum many-body physics calculations with large language models},
	Ty = {JOUR},
	Url = {https://doi.org/10.1038/s42005-025-01956-y},
	Volume = {8},
	Year = {2025}}

@misc{Cirq,
  author = {{Cirq Developers}},
  title = {{Cirq}},
  year = {2026},
  publisher = {Zenodo},
  doi = {10.5281/zenodo.4062499},
}

@misc{ReSA2026,
  author = {Michelle Barker and Daniel S. Katz and Kim Hartley and others},
  title = {{Research Software Engineering in the Age of Generative AI: Building a Community Vision}},
  howpublished = {Research Software Alliance},
  year = {2026},
  doi = {10.5281/zenodo.20320884},
}

@misc{ai4qoss,
  author = {{Unitary Foundation}},
  title = {{ai4qoss}},
  year = {2026},
  howpublished = {\url{https://github.com/unitaryfoundation/ai4qoss}},
}

@misc{unitaryHACK,
  author = {{Unitary Foundation}},
  title = {{unitaryHACK}},
  year = {2026},
  howpublished = {\url{https://unitaryhack.dev}},
  note = {accessed 2026},
}

@misc{unitaryHACK2026,
  author = {{Unitary Foundation}},
  title = {{Quantum Open Source in the Age of Coding Agents: unitaryHACK26}},
  year = {2026},
  howpublished = {\url{https://unitary.foundation/posts/2026_unitaryhack_recap/}},
}

@misc{QOSSsurvey2026,
  author = {{Unitary Foundation}},
  title = {{2026 Quantum Open Source Software Survey}},
  year = {2026},
  howpublished = {\url{https://unitary.foundation/posts/2026_qoss_survey/}},
  note = {data as of 26 October 2026; collection open through 29 October 2026},
}

@misc{ucc62,
  author = {{Unitary Foundation}},
  year = {2024},
  title = {{Verify logical equivalence of large compiled circuits, ucc issue \#62}},
  howpublished = {\url{https://github.com/unitaryfoundation/ucc/issues/62}},
  note = {opened November 2024, closed June 2025},
}

@misc{Aer2107,
  author = {{Qiskit Aer}},
  year = {2024},
  title = {{The `stabilizer' and `matrix\_product\_state' methods return incorrect results for more than 64 qubits, qiskit-aer issue \#2107}},
  howpublished = {\url{https://github.com/Qiskit/qiskit-aer/issues/2107}},
  note = {opened April 2024, fixed in release 0.14.1},
}

@misc{Stenberg2026,
  author = {Daniel Stenberg},
  title = {{The end of the curl bug-bounty}},
  year = {2026},
  howpublished = {\url{https://daniel.haxx.se/blog/2026/01/26/the-end-of-the-curl-bug-bounty/}},
}

@misc{NVQLink,
  author = {{NVIDIA}},
  title = {{NVIDIA Introduces NVQLink: Connecting Quantum and GPU Computing for 17 Quantum Builders and Nine Scientific Labs}},
  year = {2025},
  month = oct,
  howpublished = {\url{https://nvidianews.nvidia.com/news/nvidia-nvqlink-quantum-gpu-computing}},
  note = {{NVIDIA} Newsroom press release, 28 October 2025},
}

@article{Sivarajah2021,
  author = {Seyon Sivarajah and Silas Dilkes and Alexander Cowtan and Will Simmons and Alec Edgington and Ross Duncan},
  title = {{t$\vert$ket$\rangle$: a retargetable compiler for NISQ devices}},
  journal = {Quantum Sci. Technol.},
  volume = {6},
  pages = {014003},
  year = {2021},
  doi = {10.1088/2058-9565/ab8e92},
  number = {1},
}

@misc{BQSKit,
  author = {Ed Younis and Costin Iancu and Wim Lavrijsen and Marc Davis and Ethan Smith},
  title = {{Berkeley Quantum Synthesis Toolkit (BQSKit) v1}},
  year = {2021},
  howpublished = {Computer software, Lawrence Berkeley National Laboratory},
  doi = {10.11578/dc.20210603.2},
}

@article{Lambert2026,
  author = {Neill Lambert and Eric Giguère and Paul Menczel and Boxi Li and Patrick Hopf and Gerardo Suárez and Marc Gali and Jake Lishman and Rushiraj Gadhvi and Rochisha Agarwal and Asier Galicia and Nathan Shammah and Paul Nation and J.R. Johansson and Shahnawaz Ahmed and Simon Cross and Alexander Pitchford and Franco Nori},
  title = {{QuTiP 5: The Quantum Toolbox in Python}},
  journal = {Phys. Rep.},
  volume = {1153},
  pages = {1--62},
  year = {2026},
  doi = {10.1016/j.physrep.2025.10.001},
}

@article{Puzzuoli2023,
  author = {Daniel Puzzuoli and Christopher J. Wood and Daniel J. Egger and Benjamin Rosand and Kento Ueda},
  title = {{Qiskit Dynamics: A Python package for simulating the time dynamics of quantum systems}},
  journal = {J. Open Source Softw.},
  volume = {8},
  pages = {5853},
  year = {2023},
  doi = {10.21105/joss.05853},
  number = {90},
}

@article{Wittler2021,
  author = {Nicolas Wittler and Federico Roy and Kevin Pack and Max Werninghaus and Anurag Saha Roy and Daniel J. Egger and Stefan Filipp and Frank K. Wilhelm and Shai Machnes},
  title = {{Integrated Tool Set for Control, Calibration, and Characterization of Quantum Devices Applied to Superconducting Qubits}},
  journal = {Phys. Rev. Applied},
  volume = {15},
  pages = {034080},
  year = {2021},
  doi = {10.1103/PhysRevApplied.15.034080},
  number = {3},
}

@article{Goerz2019,
  author = {Michael H. Goerz and Daniel Basilewitsch and Fernando Gago-Encinas and Matthias G. Krauss and Karl P. Horn and Daniel M. Reich and Christiane P. Koch},
  title = {{Krotov: A Python implementation of Krotov's method for quantum optimal control}},
  journal = {SciPost Phys.},
  volume = {7},
  pages = {080},
  year = {2019},
  doi = {10.21468/SciPostPhys.7.6.080},
  number = {6},
}

@article{Higgott2025,
  author = {Oscar Higgott and Craig Gidney},
  title = {{Sparse Blossom: correcting a million errors per core second with minimum-weight matching}},
  journal = {Quantum},
  volume = {9},
  pages = {1600},
  year = {2025},
  doi = {10.22331/q-2025-01-20-1600},
}

@misc{VanDerWalt2026AI,
  author       = {van der Walt, St{\'e}fan},
  title        = {Community Considerations Around {AI} Contributions},
  year         = {2026},
  month        = jan,
  howpublished = {Scientific Python Blog},
  url          = {https://blog.scientific-python.org/scientific-python/community-considerations-around-ai/}
}

@misc{Hannon2026AI,
  author       = {Kevin Hannon},
  title        = {Open source maintainership in the age of {AI}},
  year         = {2026},
  month        = jun,
  howpublished = {Kubernetes Blog},
  url          = {https://kubernetes.io/blog/2026/06/26/open-source-maintainership-in-the-age-of-ai/}
}

@inproceedings{2607.29178,
  author = {Wangyiyao Zhou and Alexander Serebrenik and Alexander Nolte},
  title = {{Quick Build, Careful Check? Generative AI Use in Hackathons}},
  booktitle = {{Proceedings of the 2026 International Conference on Game Jams, Hackathons and Game Creation Events}},
  pages = {23--27},
  year = {2026},
  doi = {10.1145/3833089.3833091},
}

@inproceedings{Wu2023,
  author = {Yue Wu and Lin Zhong},
  title = {{Fusion Blossom: Fast MWPM Decoders for QEC}},
  booktitle = {{2023 IEEE International Conference on Quantum Computing and Engineering (QCE)}},
  publisher = {IEEE},
  pages = {928--938},
  year = {2023},
  doi = {10.1109/QCE57702.2023.00107},
}

@article{McClean2020,
  author = {Jarrod R McClean and Nicholas C Rubin and Kevin J Sung and Ian D Kivlichan and Xavier Bonet-Monroig and Yudong Cao and Chengyu Dai and E Schuyler Fried and Craig Gidney and Brendan Gimby and Pranav Gokhale and Thomas Häner and Tarini Hardikar and Vojtěch Havlíček and Oscar Higgott and Cupjin Huang and Josh Izaac and Zhang Jiang and Xinle Liu and Sam McArdle and Matthew Neeley and Thomas O’Brien and Bryan O’Gorman and Isil Ozfidan and Maxwell D Radin and Jhonathan Romero and Nicolas P D Sawaya and Bruno Senjean and Kanav Setia and Sukin Sim and Damian S Steiger and Mark Steudtner and Qiming Sun and Wei Sun and Daochen Wang and Fang Zhang and Ryan Babbush},
  title = {{OpenFermion: the electronic structure package for quantum computers}},
  journal = {Quantum Sci. Technol.},
  volume = {5},
  pages = {034014},
  year = {2020},
  doi = {10.1088/2058-9565/ab8ebc},
  number = {3},
}

@article{Suzuki2021,
  author = {Yasunari Suzuki and Yoshiaki Kawase and Yuya Masumura and Yuria Hiraga and Masahiro Nakadai and Jiabao Chen and Ken M. Nakanishi and Kosuke Mitarai and Ryosuke Imai and Shiro Tamiya and Takahiro Yamamoto and Tennin Yan and Toru Kawakubo and Yuya O. Nakagawa and Yohei Ibe and Youyuan Zhang and Hirotsugu Yamashita and Hikaru Yoshimura and Akihiro Hayashi and Keisuke Fujii},
  title = {{Qulacs: a fast and versatile quantum circuit simulator for research purpose}},
  journal = {Quantum},
  volume = {5},
  pages = {559},
  year = {2021},
  doi = {10.22331/q-2021-10-06-559},
}

@misc{Isakov2021,
  author = {Sergei V. Isakov and Dvir Kafri and Orion Martin and Catherine Vollgraff Heidweiller and Wojciech Mruczkiewicz and Matthew P. Harrigan and Nicholas C. Rubin and Ross Thomson and Michael Broughton and Kevin Kissell and Evan Peters and Erik Gustafson and Andy C. Y. Li and Henry Lamm and Gabriel Perdue and Alan K. Ho and Doug Strain and Sergio Boixo},
  title = {{Simulations of Quantum Circuits with Approximate Noise using qsim and Cirq}},
  year = {2021},
  eprint = {2111.02396},
  archivePrefix = {arXiv},
  primaryClass = {quant-ph},
}

@misc{Chen2021,
  author = {Mark Chen and Jerry Tworek and Heewoo Jun and Qiming Yuan and Henrique Ponde de Oliveira Pinto and Jared Kaplan and Harri Edwards and Yuri Burda and Nicholas Joseph and Greg Brockman and Alex Ray and Raul Puri and Gretchen Krueger and Michael Petrov and Heidy Khlaaf and Girish Sastry and Pamela Mishkin and Brooke Chan and Scott Gray and Nick Ryder and Mikhail Pavlov and Alethea Power and Lukasz Kaiser and Mohammad Bavarian and Clemens Winter and Philippe Tillet and Felipe Petroski Such and Dave Cummings and Matthias Plappert and Fotios Chantzis and Elizabeth Barnes and Ariel Herbert-Voss and William Hebgen Guss and Alex Nichol and Alex Paino and Nikolas Tezak and Jie Tang and Igor Babuschkin and Suchir Balaji and Shantanu Jain and William Saunders and Christopher Hesse and Andrew N. Carr and Jan Leike and Josh Achiam and Vedant Misra and Evan Morikawa and Alec Radford and Matthew Knight and Miles Brundage and Mira Murati and Katie Mayer and Peter Welinder and Bob McGrew and Dario Amodei and Sam McCandlish and Ilya Sutskever and Wojciech Zaremba},
  title = {{Evaluating Large Language Models Trained on Code}},
  year = {2021},
  eprint = {2107.03374},
  archivePrefix = {arXiv},
  primaryClass = {cs.LG},
}

@misc{Becker2025,
  author = {Joel Becker and Nate Rush and Elizabeth Barnes and David Rein},
  title = {{Measuring the Impact of Early-2025 AI on Experienced Open-Source Developer Productivity}},
  year = {2025},
  eprint = {2507.09089},
  archivePrefix = {arXiv},
  primaryClass = {cs.AI},
}

@article{Watanabe2026,
  author = {Miku Watanabe and Hao Li and Yutaro Kashiwa and Brittany Reid and Hajimu Iida and Ahmed E. Hassan},
  title = {{On the Use of Agentic Coding: An Empirical Study of Pull Requests on GitHub}},
  journal = {ACM Trans. Softw. Eng. Methodol.},
  volume = {35},
  pages = {1--24},
  year = {2026},
  doi = {10.1145/3798166},
  number = {10},
}

@article{Baltes2026,
  author = {Sebastian Baltes and Marc Cheong and Christoph Treude},
  title = {{``An Endless Stream of AI Slop'': How Developers Discuss the Burden of AI-Assisted Software Development}},
  journal = {IEEE Softw.},
  pages = {1--8},
  year = {2026},
  doi = {10.1109/MS.2026.3730626},
}

@misc{Afroz2026,
  author = {Sadia Afroz and Courtney Miller and Tyler Menezes and Edward Gilmour and Anita Sarma and Zixuan Feng},
  title = {{``AI Slop is DDoSing Open Source'': Understanding the Impact of AI-Generated Contributions on Open Source Sustainability}},
  year = {2026},
  eprint = {2607.04003},
  archivePrefix = {arXiv},
  primaryClass = {cs.SE},
}

@article{Paltenghi2022,
  author = {Matteo Paltenghi and Michael Pradel},
  title = {{Bugs in Quantum Computing Platforms: An Empirical Study}},
  journal = {Proc. ACM Program. Lang.},
  volume = {6},
  pages = {1--27},
  year = {2022},
  doi = {10.1145/3527330},
  number = {OOPSLA1},
  eid = {86},
}

@inproceedings{Paltenghi2023,
  author = {Matteo Paltenghi and Michael Pradel},
  title = {{MorphQ: Metamorphic Testing of the Qiskit Quantum Computing Platform}},
  booktitle = {{2023 IEEE/ACM 45th International Conference on Software Engineering (ICSE)}},
  publisher = {IEEE},
  pages = {2413--2424},
  year = {2023},
  doi = {10.1109/ICSE48619.2023.00202},
}

@article{Paltenghi2024,
  author = {Matteo Paltenghi and Michael Pradel},
  title = {{Analyzing Quantum Programs with LintQ: A Static Analysis Framework for Qiskit}},
  journal = {Proc. ACM Softw. Eng.},
  volume = {1},
  pages = {2144--2166},
  year = {2024},
  doi = {10.1145/3660802},
  number = {FSE},
  eid = {95},
}

@inproceedings{Wang2021,
  author = {Jiyuan Wang and Qian Zhang and Guoqing Harry Xu and Miryung Kim},
  title = {{QDiff: Differential Testing of Quantum Software Stacks}},
  booktitle = {{2021 36th IEEE/ACM International Conference on Automated Software Engineering (ASE)}},
  publisher = {IEEE},
  pages = {692--704},
  year = {2021},
  doi = {10.1109/ASE51524.2021.9678792},
}

@inproceedings{Zhao2021,
  author = {Pengzhan Zhao and Jianjun Zhao and Zhongtao Miao and Shuhan Lan},
  title = {{Bugs4Q: A Benchmark of Real Bugs for Quantum Programs}},
  booktitle = {{2021 36th IEEE/ACM International Conference on Automated Software Engineering (ASE)}},
  publisher = {IEEE},
  pages = {1373--1376},
  year = {2021},
  doi = {10.1109/ASE51524.2021.9678908},
}

@inproceedings{Lara2026,
  author = {Angel Luis Lara-Mart{\'i}n and Ricardo P{\'e}rez-Castillo},
  title = {{Entangle: Uncovering Collaboration in the GitHub Quantum Software Ecosystem}},
  booktitle = {{2026 IEEE International Conference on Quantum Computing and Engineering (QCE)}},
  publisher = {IEEE},
  year = {2026},
  eprint = {2609.28349},
  archivePrefix = {arXiv},
  primaryClass = {cs.SE},
}

@misc{2603.08801,
  author = {Shiheng Li and Jacob M. Miller and Phoebe J. Lee and Gustav Andersson and Christopher R. Conner and Yash J. Joshi and Bayan Karimi and Amber M. King and Howard L. Malc and Harsh Mishra and Hong Qiao and Minseok Ryu and Xuntao Wu and Siyuan Xing and Haoxiong Yan and Jian Shi and Andrew N. Cleland},
  title = {{Large Language Model-Assisted Superconducting Qubit Experiments}},
  year = {2026},
  eprint = {2603.08801},
  archivePrefix = {arXiv},
  primaryClass = {quant-ph},
}

@misc{2606.22376,
  author = {Huikai Xu and Jiaxiu Han and Shigang Ou and Cheng Ye and Zisong Shen and Jing Gao and Yijia Wang and Tianrui Che and Yu Song and Weiyang Liu and Lei Wang and Lin-Feng Zhang and Pan Zhang and Hai-Feng Yu},
  title = {{Vibe Calibration: Autonomous Bring-up of a 112-Qubit Superconducting Quantum Processor by a Skill-Orchestrating Language Agent}},
  year = {2026},
  eprint = {2606.22376},
  archivePrefix = {arXiv},
  primaryClass = {quant-ph},
}

@misc{2607.25145,
  author = {Takuya Isogawa and Ryotaro Okabe and Nutdech Phadetsuwannukun and Mingda Li and Paola Cappellaro},
  title = {{Agentic AI for Scientific Reasoning in Autonomous Quantum Sensing Experiments}},
  year = {2026},
  eprint = {2607.25145},
  archivePrefix = {arXiv},
  primaryClass = {quant-ph},
}

@misc{2609.17439,
  author = {Naixu Guo and Changhao Li and Siyu Cheng and Qicheng Tang and Binzhao Luo and Bikun Li and Yuxuan Du and Shihao Ru and Jiaqi Cai},
  title = {{Evaluating Verified Autonomy in Quantum Engineering}},
  year = {2026},
  eprint = {2609.17439},
  archivePrefix = {arXiv},
  primaryClass = {quant-ph},
}

@misc{qiqcop_zoo,
  title = {{Quantum Information and Quantum Computation Open Problem Zoo (QIQCOP Zoo)}},
  howpublished = {\url{https://qiqc-op.com/}},
  year = {2026}
}

@article{2305.06378,
  author = {Vincent Paul Su and ChunJun Cao and Hong-Ye Hu and Yariv Yanay and Charles Tahan and Brian Swingle},
  title = {{Discovery of optimal quantum codes via reinforcement learning}},
  journal = {Phys. Rev. Applied},
  volume = {23},
  pages = {034048},
  year = {2025},
  doi = {10.1103/PhysRevApplied.23.034048},
  number = {3},
}

@article{Olle2024,
  author = {Jan Olle and Remmy Zen and Matteo Puviani and Florian Marquardt},
  title = {{Simultaneous discovery of quantum error correction codes and encoders with a noise-aware reinforcement learning agent}},
  journal = {npj Quantum Inf.},
  volume = {10},
  pages = {126},
  year = {2024},
  doi = {10.1038/s41534-024-00920-y},
  number = {1},
}

@misc{2502.14372,
  author = {Austin Yubo He and Zi-Wen Liu},
  title = {{Discovering highly efficient low-weight quantum error-correcting codes with reinforcement learning}},
  year = {2025},
  eprint = {2502.14372},
  archivePrefix = {arXiv},
  primaryClass = {quant-ph},
}

@misc{2607.25865,
  author = {Ge Yan and Shanchuan Li and Pengyue Ma and Qixin Zhang and Pingchuan Ma and Jianping Wang and Min-Hsiu Hsieh and Yuxuan Du},
  title = {{OmniQEC: discovering practical quantum error-correcting codes by an AI scientist}},
  year = {2026},
  eprint = {2607.25865},
  archivePrefix = {arXiv},
  primaryClass = {quant-ph},
}

@misc{QECChallenge2026,
  author = {{Unitary Foundation}},
  title = {{An open leaderboard for human- and AI-discovered quantum codes}},
  year = {2026},
  howpublished = {Unitary Foundation, \url{https://unitary.foundation/posts/2026_qec_challenge/}},
}

@inproceedings{Xia2026KQFuzz,
  author = {Fuyuan Xia and Qixin Zhang and Chenhao Ying and Haojin Zhu and Shuai Wang and Yuan Luo and Pingchuan Ma and Yuxuan Du},
  title = {{KQFuzz: Knowledge-Guided Fuzzing for Quantum Libraries via Large Language Models}},
  booktitle = {Proceedings of the 41st IEEE/ACM International Conference on Automated Software Engineering},
  year = {2026},
  doi = {10.1145/3832783.3834392},
  pages = {777--789},
  publisher = {ACM},
}

@misc{2607.25834,
  author = {Constantin Dalyac and Alexandre Dauphin and Lo{\"i}c Henriet and Christophe Jurczak},
  title = {{Lowering the implementation barrier of neutral-atom quantum computing with agentic workflows}},
  year = {2026},
  eprint = {2607.25834},
  archivePrefix = {arXiv},
  primaryClass = {quant-ph},
}

@misc{qBraidCodeq2026,
  author = {{qBraid}},
  title = {{qBraid for Developers: Build Quantum Applications}},
  year = {2026},
  howpublished = {\url{https://qbraid.com/developers}},
  note = {accessed 22 October 2026},
}

@article{Ladd2010,
  author = {T. D. Ladd and F. Jelezko and R. Laflamme and Y. Nakamura and C. Monroe and J. L. O'Brien},
  title = {{Quantum computers}},
  journal = {Nature},
  volume = {464},
  pages = {45--53},
  year = {2010},
  doi = {10.1038/nature08812},
  number = {7285},
}

@article{Georgescu2014,
  author = {I. M. Georgescu and S. Ashhab and Franco Nori},
  title = {{Quantum simulation}},
  journal = {Rev. Mod. Phys.},
  volume = {86},
  pages = {153--185},
  year = {2014},
  doi = {10.1103/RevModPhys.86.153},
  number = {1},
}

@article{Degen2017,
  author = {C. L. Degen and F. Reinhard and P. Cappellaro},
  title = {{Quantum sensing}},
  journal = {Rev. Mod. Phys.},
  volume = {89},
  pages = {035002},
  year = {2017},
  doi = {10.1103/RevModPhys.89.035002},
  number = {3},
}

@article{Wehner2018,
  author = {Stephanie Wehner and David Elkouss and Ronald Hanson},
  title = {{Quantum internet: A vision for the road ahead}},
  journal = {Science},
  volume = {362},
  pages = {eaam9288},
  year = {2018},
  doi = {10.1126/science.aam9288},
  number = {6412},
}

@misc{2606.13811,
  author = {Rogerio Feris and Yunchao Liu and Pengyuan Li and Hang Hua and David Kremer},
  title = {{Aligning Quantum Operators with Large Language Models}},
  year = {2026},
  eprint = {2606.13811},
  archivePrefix = {arXiv},
  primaryClass = {quant-ph},
}

@misc{2607.24714,
  author = {Fabian Kreppel and Reza Salkhordeh and Ferdinand Schmidt-Kaler and Andr{\'e} Brinkmann},
  title = {{Efficient LLM-Generated Shuttling Compilers for Complex Trapped-Ion Architectures}},
  year = {2026},
  eprint = {2607.24714},
  archivePrefix = {arXiv},
  primaryClass = {quant-ph},
}

@misc{2609.38883,
  author = {Lily Jiaxin Wan and Deming Chen and Klara Nahrstedt and Bo Chen},
  title = {{Toward Quantum Software Automation: A Quantum-Aware Harness for LLM-Guided Evolution}},
  year = {2026},
  eprint = {2609.38883},
  archivePrefix = {arXiv},
  primaryClass = {cs.SE},
}

@misc{2610.22367,
  author = {Andi Gu and Alek H. Kemeny and Susanne F. Yelin and Mikhail D. Lukin},
  title = {{Discovering state-of-the-art quantum circuits with large language models}},
  year = {2026},
  eprint = {2610.22367},
  archivePrefix = {arXiv},
  primaryClass = {quant-ph},
}

@misc{2606.27231,
  author = {Duanyang Wang and Lu Qi and Yuanheng Xie and Norbert M. Linke and Kenneth R. Brown},
  title = {{A hardware-safety-gated system for LLM-written native ARTIQ control code on a trapped-ion platform}},
  year = {2026},
  eprint = {2606.27231},
  archivePrefix = {arXiv},
  primaryClass = {quant-ph},
}

@misc{2604.25884,
  author = {Shuxiang Cao and Zijian Zhang and Abhishek Agarwal and Grace Bratrud and Niyaz R. Beysengulov and Daniel C. Cole and Alejandro G{\'o}mez Frieiro and Elena O. Glen and Hao Hsu and Gang Huang and Raymond Jow and Greshma Shaji and Tom Lubowe and Ligeng Zhu and Luis Mantilla Calder{\'o}n and Nicola Pancotti and Joel Pendleton and Brandon Severin and Charles Etienne Staub and Sara Sussman and Antti Veps{\"a}l{\"a}inen and Neel Rajeshbhai Vora and Yilun Xu and Varinia Bernales and Daniel Bowring and Elica Kyoseva and Ivan Rungger and Giulia Semeghini and Sam Stanwyck and Timothy Costa and Al{\'a}n Aspuru-Guzik and Krysta Svore},
  title = {{QCalEval: Benchmarking Vision-Language Models for Quantum Calibration Plot Understanding}},
  year = {2026},
  eprint = {2604.25884},
  archivePrefix = {arXiv},
  primaryClass = {quant-ph},
}

@misc{2606.02418,
  author = {Juan Cruz-Benito and Andrew W. Cross and David Kremer and Ismael Faro},
  title = {{Evolutionary Discovery of Bivariate Bicycle Codes with LLM-Guided Search}},
  year = {2026},
  eprint = {2606.02418},
  archivePrefix = {arXiv},
  primaryClass = {quant-ph},
}

@misc{2606.24808,
  author = {Zidu Liu and Florian Marquardt},
  title = {{Large-Language-Model Discovery of Quantum LDPC Codes through Structured Concept Evolution}},
  year = {2026},
  eprint = {2606.24808},
  archivePrefix = {arXiv},
  primaryClass = {quant-ph},
}

@misc{2510.20728,
  author = {Xi He and Sirui Lu and Bei Zeng},
  title = {{Co-Designing Quantum Codes with Transversal Diagonal Gates via Multi-Agent Systems}},
  year = {2025},
  eprint = {2510.20728v3},
  archivePrefix = {arXiv},
  primaryClass = {quant-ph},
  note = {Version 3, revised April 2026},
}

@misc{2604.05587,
  author = {Zhe Zhao and Haibin Wen and Jiaming Ma and Jiachang Zhan and Tianyi Xu and Ye Wei and Qingfu Zhang},
  title = {{ResearchEVO: An End-to-End Framework for Automated Scientific Discovery and Documentation}},
  year = {2026},
  eprint = {2604.05587},
  archivePrefix = {arXiv},
  primaryClass = {cs.AI},
}

@misc{Bhandwaldar2026,
  author = {Abhishek Bhandwaldar and Mihir Choudhury and Amelia Lopez and Carlos H. A. Costa and Gi-Joon Nam and David S. Kung and Ruchir Puri},
  title = {{Highly Cost-Efficient Agent Factories for High Level Synthesis: Design Space Exploration for Quantum Error Correction Chip with Open Frontier Models and Distributed Inference at Scale}},
  year = {2026},
  howpublished = {IBM Research preprint},
  url = {https://www.ibm.com/downloads/documents/us-en/17f55bd44d44b0e3},
  note = {accessed 22 October 2026},
}

@misc{2307.08191,
  author = {Zhiding Liang and Jinglei Cheng and Rui Yang and Hang Ren and Zhixin Song and Di Wu and Xuehai Qian and Tongyang Li and Yiyu Shi},
  title = {{Unleashing the Potential of LLMs for Quantum Computing: A Study in Quantum Architecture Design}},
  year = {2023},
  eprint = {2307.08191},
  archivePrefix = {arXiv},
  primaryClass = {quant-ph},
}

@inproceedings{2509.08385,
  author = {Yaswitha Gujju and Romain Harang and Tetsuo Shibuya},
  title = {{LLM-Guided Ans{\"a}tze Design for Quantum Circuit Born Machines in Financial Generative Modeling}},
  year = {2025},
  note = {Presented at QML@QCE 2025},
  doi = {10.1109/qce65121.2025.10303},
  booktitle = {2025 IEEE International Conference on Quantum Computing and Engineering (QCE)},
  pages = {104--108},
  publisher = {IEEE},
}

@article{Peng2023Shor,
  author = {Yuxiang Peng and Kesha Hietala and Runzhou Tao and Liyi Li and Robert Rand and Michael Hicks and Xiaodi Wu},
  title = {{A formally certified end-to-end implementation of Shor's factorization algorithm}},
  journal = {Proc. Natl. Acad. Sci. U.S.A.},
  volume = {120},
  number = {21},
  pages = {e2218775120},
  year = {2023},
  doi = {10.1073/pnas.2218775120},
}

@article{Hietala2021VOQC,
  author = {Kesha Hietala and Robert Rand and Shih-Han Hung and Xiaodi Wu and Michael Hicks},
  title = {{A verified optimizer for quantum circuits}},
  journal = {Proc. ACM Program. Lang.},
  volume = {5},
  number = {POPL},
  pages = {1--29},
  year = {2021},
  doi = {10.1145/3434318},
}

@article{Zhou2023CoqQ,
  author = {Li Zhou and Gilles Barthe and Pierre-Yves Strub and Junyi Liu and Mingsheng Ying},
  title = {{CoqQ: Foundational Verification of Quantum Programs}},
  journal = {Proc. ACM Program. Lang.},
  volume = {7},
  number = {POPL},
  pages = {833--865},
  year = {2023},
  doi = {10.1145/3571222},
}

@article{Arlt2026,
  author = {S{\"o}ren Arlt and Haonan Duan and Felix Li and Sang Michael Xie and Yuhuai Wu and Mario Krenn},
  title = {{Meta-designing quantum experiments with language models}},
  journal = {Nat. Mach. Intell.},
  volume = {8},
  pages = {148--157},
  year = {2026},
  doi = {10.1038/s42256-025-01153-0},
  number = {2},
}

@misc{2610.23832,
  title = {Leveraging verifiable structure in quantum circuits to find more efficient compilations with agentic {AI}},
  author = {Tanuj Khattar and Nicholas C. Rubin and Craig Gidney and Michael Broughton and Guang Hao Low and Ryan Babbush},
  year = {2026},
  eprint = {2610.23832},
  archivePrefix = {arXiv},
  primaryClass = {quant-ph},
}

@misc{UnitaryAgenticLeaderboards2026,
  author = {{Unitary Foundation}},
  title  = {Agentic leaderboards for quantum information research},
  year   = {2026},
  month  = oct,
  howpublished = {\url{https://unitary.foundation/posts/2026_agentic_leaderboards/}},
}

@misc{StabilizerRankChallenge2026,
  author = {{Unitary Foundation}},
  title  = {Stabilizer Rank Challenge},
  year   = {2026},
  howpublished = {\url{https://unitaryfoundation.github.io/stabrank/}},
}

@misc{ECDSAFail2026,
  author = {{Eigen Labs}},
  title  = {{ECDSA.fail}: Can you break {ECDSA}?},
  year   = {2026},
  howpublished = {\url{https://www.ecdsa.fail/}},
}

@misc{MetriqGymBarrierFix2026,
  author = {{Unitary Foundation}},
  title  = {Require {amazon-braket-sdk}$\geq$1.111.1 to preserve barriers},
  year   = {2026},
  howpublished = {\url{https://github.com/unitaryfoundation/metriq-gym/pull/783}},
  note   = {metriq-gym pull request \#783.}
}

@misc{2603.08680,
  title = {Metriq: A Collaborative Platform for Benchmarking Quantum Computers},
  author = {Alessandro Cosentino and Changhao Li and Vincent Russo and Bradley A. Chase and Tom Lubinski and Siyuan Niu and Neer Patel and Nathan Shammah and William J. Zeng},
  year = {2026},
  eprint = {2603.08680},
  archivePrefix = {arXiv},
  primaryClass = {quant-ph},
}

@misc{2608.28433,
  title = {{Prove2Me}: An Open Collaborative Platform for Scaling Math Formalization},
  author = {Shuze Chen and Kunal Marwaha and Xiaoyang Lu and Henry Yuen and Tianyi Peng},
  year = {2026},
  eprint = {2608.28433},
  archivePrefix = {arXiv},
  primaryClass = {cs.AI},
}

\end{document}